\documentclass{imsart}
\pubyear{2024}
\volume{00}
\issue{0}
\doi{0000}
\firstpage{1}
\lastpage{1}

\usepackage{amsmath}
\usepackage{graphicx}
\usepackage{enumerate}
\usepackage{natbib}
\usepackage{url} 
\usepackage{placeins}

\usepackage{amsfonts}
\usepackage{amssymb}
\usepackage{amsmath}
\usepackage{amsthm}
\usepackage{pdflscape}
\usepackage{latexsym,amsmath}
\usepackage{multirow}
\usepackage{caption}
\usepackage{tikz}
\usepackage{hyperref}
\usepackage{afterpage}
\usepackage{soul}
\usepackage{xcolor}
\usepackage{xurl}
\usepackage{subfig}

\newtheorem{theorem}{Theorem}

\begin{document}

\begin{frontmatter}
\title{Bayesian Consistency for Long Memory Processes: A Semiparametric Perspective}
\runtitle{Semiparametric Bayesian Model for Long Memory Processes}

\begin{aug}
\author{\fnms{Clara} \snm{Grazian}\thanksref{addr1,addr2}\ead[label=e1]{clara.grazian@sydney.edu.au}},

\runauthor{Grazian, C.}

\address[addr1]{Carslaw Building, University of Sydney, Camperdown Campus, Sydney, Australia
    \printead{e1} 
}
\address[addr2]{ARC Training Centre in Data Analytics for Resources and Environments (DARE), Sydney, Australia
}
%

\end{aug}

\begin{abstract}
In this work, we will investigate a Bayesian approach to estimating the parameters of long memory models. Long memory, characterized by the phenomenon of hyperbolic autocorrelation decay in time series, has garnered significant attention. This is because, in many situations, the assumption of short memory, such as the Markovianity assumption, can be deemed too restrictive. Applications for long memory models can be readily found in fields such as astronomy, finance, and environmental sciences. However, current parametric and semiparametric approaches to modeling long memory present challenges, particularly in the estimation process.

In this study, we will introduce various methods applied to this problem from a Bayesian perspective, along with a novel semiparametric approach for deriving the posterior distribution of the long memory parameter. Additionally, we will establish the asymptotic properties of the model. An advantage of this approach is that it allows to implement state-of-the-art efficient algorithms for nonparametric Bayesian models. 
\end{abstract}

\noindent%
{\it Keywords:}  non-Gaussian long memory processes, dependent Dirichlet processes, spectral density, log-periodogram
\vfill

\end{frontmatter}

\section{Introduction}
\label{sec:intro}

Over the past few decades, there has been significant growth in both theoretical and applied literature concerning the phenomenon of long memory. This phenomenon refers to the slow decay of correlations between observations that are increasingly distant from each other, maintaining non-negligible levels of correlation. Long memory presents an alternative correlation structure to that found in the popular ARMA (autoregressive moving average) model and its differenced extensions, the ARIMA (autoregressive integrated moving average) model. These models, which are standard tools in time series analysis, exhibit short memory, meaning their correlations decay geometrically to zero.

Literature has burgeoned from a wide array of scientific fields where observed data appears to exhibit long memory. These fields include economics and finance \citep{ASIF2022126871}, hydrology \citep{Pelletier1997LongrangePI}, astronomy \citep{Feigelson2018AutoregressiveTS}, climate science \citep{climate_paper}, and medical science \citep{refId0}. Alongside literature on the application of long memory models, there has been ongoing research into methods of detecting and performing inference in the presence of long memory in a series; see \cite{palma2007long,beran_longmem,hassler2018time} for recent reviews.

Consider a univariate stochastic process $\{x_t\}_{t \in \mathcal{T}}$, which is a second-order stochastic process with $\mathbb{E}(x_t)=\mu$ and $Cov(x_t, x_{t+h}) = \gamma(h)$ for all $t, t+h \in \mathcal{T}$. The function $\gamma(h)$ is called the autocovariance of the process. Consider then the harmonic frequencies of a sample $x_1, \ldots, x_n$, also called Fourier frequencies, defined as $
\lambda_j = \frac{2\pi j}{n}$ for $j=1, \ldots, \lfloor \frac{n}{2}\rfloor$, where the symbol $\lfloor a \rfloor$ denotes the largest integer not greater than $a$. It is enough to consider only $j \geq 1$ and $j\leq \lfloor \frac{n}{2}\rfloor$ because a series of a longer period (or smaller frequency than $\lambda_1$) cannot be observed from a sample of length $n$. Frequencies larger than $\pi$ are not considered because they would correspond to a period shorter than two, which is not observable in discrete time. The time series can then be expressed as a combination of cosine and sine waves, with coefficients given by the Fast Fourier Transform (FFT). The periodogram is defined in relation to estimates of these coefficients:
$$
I(\lambda_j) = \frac{1}{2\pi n} \left[ \left( \sum_{t=1}^n x_t \cos(\lambda_j t) \right)^2 + \left( \sum_{t=1}^n x_t \sin(\lambda_j t) \right)^2 \right].
$$
The periodogram is a sample estimate of a population function called the spectral density $f(\lambda)$, which is a frequency domain characterization of a stationary time series. However, it is not a consistent estimator, so windowing methods have been proposed to obtain a consistent estimator \citep{parzen1962estimation}.

\begin{figure}[h]
\begin{center}
\includegraphics[width=4.5in]{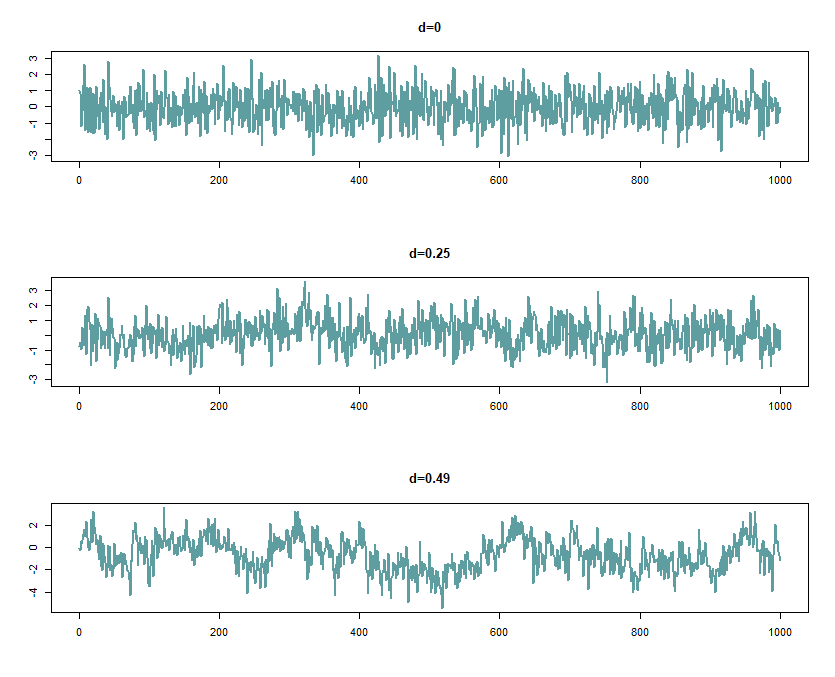}
\end{center}
\caption{Examples of time series for different values of $d$. \label{fig:longmem_ex}}
\end{figure}

A stationary process $\{x_t\}_{t \in \mathcal{T}}$ is said to have long memory \citep{giraitis2012large} if the sequence of autocovariances is not summable: $
\sum_{h=0}^{H} |\gamma(h)| \rightarrow \infty$ as $H \rightarrow \infty$; otherwise, it is said to have short memory. This corresponds to a spectral density such that
$
\lim_{\lambda \rightarrow 0} \frac{f(\lambda)}{c \lambda^{-2d}} = 1
$, for some $c \in \mathbb{R}$; the parameter $-1 < d < 1/2$ is defined as the long memory parameter. This is equivalent to saying that $
\lim_{h \rightarrow \infty} \frac{\gamma(h)}{g h^{2d-1}} = 1
$, for some $g \in \mathbb{R}$; this equivalence comes from the theory of Fourier series \citep{zygmund1959preservation}. Examples of time series for different values of $d$ are shown in Figure \ref{fig:longmem_ex}: as the long memory parameter increases, the time series displays reversing trends; as $d \rightarrow 1/2$, the series displays long time spans of growth and decrease. For $d > 1/2$, the process is non-stationary; for $-1/2 <d < 0$, the process is antipersistent, i.e., the process crosses its expectation often. The case $-1<d<-1/2$ occurs when differencing a stationary process integrated of order $\delta$, with $0 < \delta < 1/2$ \citep{hassler2018time}. 

For Gaussian parametric stationary processes characterized by long memory, \cite{dahlhaus1989efficient} and \cite{lieberman2012asymptotic} proved the asymptotic normality of the maximum likelihood estimators. However, exact methods can be computationally expensive. Several approximations exist that are computationally cheaper at the cost of some accuracy loss \citep{breidt1998detection, sela2008computationally, davis2011comments, jeong2013wavelet}. Moreover, for long memory processes, the sample mean of the process converges to the true mean at a slower rate than the usual $\sqrt{n}$, with a rate that becomes slower as $d$ increases \citep{samarov1988efficiency}. Therefore, when the mean of the process must be estimated as well, the maximum likelihood estimator tends to show strong bias in practical situations. A possible solution is to assume a parametric model, such as an ARFIMA$(p,d,q)$ model, where the lag orders $p$ and $q$ can be determined with model selection techniques, such as information criteria. For short memory processes, \cite{hannan1980estimation} shows that BIC is consistent; on the other hand, \cite{baillie2014modified} show through simulations that both AIC and BIC tend to overestimate the order of the ARMA part in the presence of long memory. Finally, the assumption of Gaussianity is a strong limitation.

Estimation of the parameter of long memory processes is also performed in the frequency domain, with the advantage that the periodogram is invariant with respect to additive constants (like the mean of the process) as long as it is evaluated at the harmonic frequencies. A possible solution is to use the approximation provided by the Whittle likelihood \citep{whittle1953estimation,baum2020local}, which has the advantage that its computation is fast thanks to the Fast Fourier Transform algorithm. Whittle estimation under long memory was first introduced by \cite{fox1986large}, and the limiting normality of the maximum likelihood estimator for the Whittle likelihood for stationary Gaussian long memory processes was proved by \cite{fox1986large} and \cite{dahlhaus1989efficient}, and by \cite{giraitis1990central} for stationary linear processes with independent and identically distributed (i.i.d.) errors. \cite{sykulski2016biased} and \cite{kirch2019beyond} have recently proposed corrections to the Whittle likelihood to remove biases: the Whittle likelihood assumes that $\frac{I(\lambda_j)}{f(\lambda_j)}$ are approximately independent and identically distributed, which is correct under the assumption of short memory, but not under long memory at low frequencies \citep{robinson_logperiog}, and the approximation is exact only in the case of Gaussian time series. Several methods have also been proposed for directly estimating $d$. \cite{fox1986large} presents a parametric method for Gaussian long-memory processes, which is generalized by \cite{GIRAITIS1999} for non-Gaussian time series. \cite{geweke1983estimation} propose a regression model for $f(\lambda)$, and \cite{beran1993fitting} use a generalized linear regression. Semiparametric methods have been proposed by \cite{giraitis1990central}, \cite{robinson_logperiog}, and \cite{robinson1995gaussian}, among others. 

Approaches to the estimation of spectral densities are available in the Bayesian literature as well. Several authors have introduced parametric approaches, particularly for ARFIMA models: see \cite{hosking_fracdiff}, \cite{ravishanker1997bayesian}, \cite{Pai1998BayesianAO}, and \cite{ko2006bayesian}. \cite{KOOP1997149} propose a fully Bayesian approach. The advantage of a fully Bayesian approach is that the posterior distributions of the parameters, including the long memory parameter $d$, are defined for finite samples and do not rely on the asymptotic normality of the estimators; therefore, they can be skewed and have fat tails. For short memory processes, \cite{carter1997semiparametric} use a mixture of normal distributions and give $\log f(\lambda)$ a smoothing prior. \cite{choudhuri2004bayesian} use Bernstein polynomials, while \cite{rosen2007automatic} define a model $\log f(\lambda) = \alpha_0 + \alpha_1 \lambda + h(\lambda)$ and a Gaussian process prior for $h(\lambda)$. Additionally, \cite{pensky2007bayesian} use Bayesian wavelet smoothing of the periodogram.

To estimate spectral density in absence of long memory,  \cite{kirch2019beyond} propose a semi-parametric approach for bootstrapping, where a parametric model acts as a proxy for the rough shape of the autocorrelation structure and the dependency structure between periodogram ordinates. Then, a nonparametric correction is applied to reduce dependence on the parametric assumptions. \cite{cadonna2019bayesian} propose a Bayesian model for the spectral density where a translated log-periodogram is described by a mixture of Gaussian distributions with frequency-dependent weights and mean functions, which converges in $L_p$-norm to the true log-spectral density and is asymptotically equivalent to the Whittle likelihood.

Here, we propose a flexible Bayesian semiparametric approach to estimating the parameter of long memory $d$, while simultaneously reducing assumptions on the distribution of the time series. For building this new approach for a univariate time series, we employ a regression approach akin to \cite{robinson_logperiog}. However, utilizing a Bayesian framework allows us to derive a posterior distribution for $d$, rather than solely relying on the asymptotic normality of the least square estimator. Furthermore, it enables us to obtain a posterior process for the distribution of the error terms of the regression, resulting in convergence results and more stable estimation of $d$.

Similar to \cite{cadonna2019bayesian}, we assume that the log-periodogram distribution follows a mixture of Gaussian distributions with frequency-dependent mixture weights. However, our method can handle processes in the presence of long memory, as well as antipersistence. A significant advantage lies in computational efficiency, as the model can leverage well-established simulation methods to perform inference for dependent Dirichlet process mixture models. This allows the proposed model to be applied to relatively large time series datasets.

The method is applied to a series of simulated and realistic examples. In particular, long memory has recently found application in areas such as finance and climate change, making these domains crucial for validating our methodology. A semiparametric Bayesian approach offers a means to ascertain the presence of long memory in real-world settings, with potential implications for policy decisions. For instance, detecting long memory in global temperature time series could suggest that policy implementations require more time to yield significant effects.

The outline of the article is as follows: in Section \ref{sec:bayesian}, we review related approaches in Bayesian semiparametric analysis of long memory processes. Section \ref{sec:meth}, details our modeling approach and presents theoretical results of consistency, with technical details provided in the appendices. Section \ref{sec:simu} presents results from an extensive simulation study. Section \ref{sec:data} applies the proposed model to three distinct datasets: a known benchmark in long memory analysis, a finance example, and a global temperature dataset. Finally, Section \ref{sec:conc} offers a summary and discusses potential extensions of the methodology.

\section{Bayesian methods to estimate spectral densities and long memory processes}
\label{sec:bayesian}

Consider a sequence of statistical models $\mathcal{F}^{(n)} = \{F_{d,\eta}^{(n)}: d \in (-1,1/2), \eta \in \mathcal{H}\}$, parameterized by a finite-dimensional parameter $d$ of interest and an infinite-dimensional parameter $\eta$, which is considered a nuisance parameter. We assume that $\mathcal{F}^{(n)}$ has density $f_{d,\eta}^{(n)}$. Let $x^{(n)}$ be a random element which follows $F_0^{(n)}$, and assume that $F_{0}^{(n)} = F_{d_0,\eta_0}^{(n)}$ for some $d_0 \in (-1, 1/2)$ and $\eta_0 \in \mathcal{H}$.

We consider a product prior $\Pi_{(-1,1/2)} \times \Pi_{\mathcal{H}}$ and denote the posterior distribution by $\Pi(\cdot | x^{(n)})$. The prior $\Pi_{(-1,1/2)}$ is assumed to be thick at $d_0$, i.e., to have a positive continuous Lebesgue density in a neighborhood of $d_0$, which is not a strong assumption, given that the space of $d$ is bounded.

In the setting of long memory processes, \cite{hurvich1993asymptotics} propose a representation of the spectral density given by
\begin{equation}
f(\lambda) = f^*(\lambda) \cdot |1-e^{-i \lambda}|^{-2d}
\label{eq:lm_hb}
\end{equation}
where $d \in (-1,1)$, $\lambda \in [0,\pi]$, and $f^*$ is a positive even continuous function on $[0,\pi]$ that is bounded away from zero on $(-\pi,\pi)$. In this representation, $d$ governs the long memory behavior, while $f^*$ governs the short memory behavior. The factor $|1-e^{-i \lambda}|$ is the frequency domain representation of the long memory component: as $\lambda \rightarrow 0$, this term approaches one, indicating that the term is a power-law decay function that reduces the power of the spectral density at higher frequencies. In what follows, we will make use of the relationship $|1-e^{-i \lambda}| = 2 \sin \left(\frac{\lambda}{2}\right)$ \citep{palma2007long}.

\cite{petris1997bayesian} use another representation of the spectral density:
\begin{equation}
f(\lambda) = \exp(g(\lambda)) \cdot \lambda^{-2d},
\label{eq:petris}
\end{equation}
where $g(\cdot)$ is a function in $C([0,\pi])$, the vector space of all continuous real-valued functions on the interval $[0,\pi]$, that captures the short memory behavior, and $d \in [0,1)$. \cite{liseo2001bayesian} show that $g(\cdot)$ needs some additional assumptions for the model to be identifiable. In particular, $\lim_{\lambda \rightarrow 0} \exp(g(\lambda)) < C$ with $0 < C < \infty$, which automatically holds if $g(\lambda)$ is continuous with probability one. If this assumption is not imposed, any short-memory dependent process with spectral density $h(\cdot)$ can produce positive values of $d$ by taking $g(\lambda) = \log[\lambda^{2d} h(\lambda)]$.

Representation \eqref{eq:petris} allows for a regression-type model, relating the periodogram to the long-memory parameter:
$$
y_{j} = \log(I(\lambda_j)) = -d \log \lambda_j + g(\lambda_j) + u_j \qquad j=1, \ldots, \lfloor n/2\rfloor.
$$
Following \cite{carter1997semiparametric}, \cite{petris1997bayesian} suggests approximating the distribution of the log-periodogram with a finite Gaussian mixture model with five components. Then
$$
y_{jk}|s_j \sim N(-d \log \lambda_j + g_j + \mu_{k}, V_{k}) \qquad k=1, \ldots, 5 \quad \text{ and } \quad j=1, \ldots, n. 
$$
The variable $s_j$ is an allocation variable, which assigns the observation given by the log-periodogram at frequency $\lambda_j$ to a specific component of the mixture model. \cite{petris1997bayesian} assume $g \perp d$, and define a Gaussian process prior for $g$ and a spike-and-slab prior for $d$. The use of a finite mixture model may represent a strong assumption and not converge to the continuous distribution of the error terms $u_j$. Additionally, \cite{petris1997bayesian}, similarly to the Whittle likelihood, assumes independence of the errors, which is not assured in the case of long memory processes.

\cite{liseo2001bayesian} make a stronger assumption on the spectral density:
$$
f(\lambda) = G \cdot \lambda^{-2d} 
$$
where $0 < G < \infty$ and $\lambda < \lambda^*$. This definition represents an approximation of Equation \eqref{eq:petris} under regularity conditions \citep{Velasco2000WhittlePL}. Beyond $\lambda^*$, the spectral density $h(\lambda)$ is modeled nonparametrically, using a vague prior distribution on $d$ and a noninformative process on the spectral density away from zero. The parameter $\lambda^*$ takes values $\lambda^* = \frac{2\pi j}{n}$, with $j={1,2,\ldots,\lfloor \frac{n}{2} \rfloor}$. The parameters of the model are $\{d,G,h(\lambda)\}$, and they are defined in $[0,1/2) \times [0,\infty) \times \mathcal{H}$, where $\mathcal{H}$ is a subspace of $L^1[\lambda^*,\pi]$, the space of integrable functions on $[\lambda^*,\pi]$. This restriction induces a model selection problem on the choice of $\lambda^*$. Moreover, given some assumptions on the model, \cite{liseo2001bayesian} derive that the prior process for the nonparametric part of the model is a Brownian motion. While flexible, this method can still be restrictive depending on the specific characteristics of the model.

\cite{Holan2009ABA} propose a semiparametric approach for fractional exponential model (FEXP) processes. The spectral density can be represented as
$$
f(\lambda) = \exp\left[\sum_{k=0}^m a_k \cos(k\lambda)\right] \cdot |1-e^{-i\lambda}|^{-2d}
$$
where $a_0, \ldots, a_m$ are constants given diffuse prior distributions, $m \in \mathbb{Z}^+$ is adaptively selected, and $d$ is given a flat prior. The parameter vector is $(d, a_0, \ldots, a_m)$, defined on $[0, 1/2) \times \mathbb{R}^{m+1}$, and depends on $m$. The coefficients $a_k$ are given a spike-and-slab prior, with a normal component for $k \leq m$ and a mass at zero for $k > m$. The long memory parameter $d$ is given a $\text{Unif}(0, 1/2)$ prior distribution to ensure the process is stationary. The order $m$ is given a discrete uniform distribution over ${1, \ldots, M}$, for some choice $M$ of the maximum value for $m$, or a truncated Poisson distribution. \cite{Holan2009ABA} propose a reversible-jump MCMC to estimate the posterior distribution of this model, which is suitable for the variable-dimension approach given the randomness of $m$.

The approach of \cite{Holan2009ABA} relies on an assumption of Gaussianity of the process, which is advantageous if the assumption is correct. For example, it allows for more natural handling of missing data and permits proving convergence of the posterior expected value to the true value $d_0$, when $d$ is independent of $m$. However, this is true only if the generating process is correctly specified. Our approach does not rely on this assumption and can be seen as a more robust approach with respect to misspecification.

The assumptions that $d \in [0,1)$ as in \cite{petris1997bayesian} or $d \in [0,1/2)$ as in \cite{liseo2001bayesian} and \cite{Holan2009ABA} are often made in works on processes with long memory. However, this implies a different treatment of processes with long memory and processes with antipersistence. We avoid this distinction in this work, and present a unified theory to estimate $d$ without an assumption that long memory is present in the observed time series. 
 
\section{Methods}
\label{sec:meth}

\cite{geweke1983estimation} propose a regression model for the spectral density at the origin, defining $f^*$ in \eqref{eq:lm_hb} as the spectral density of white noise $f_u(\lambda)$. Then it is possible to redefine the equation in logarithmic scale and relate this equation to the log-periodogram:
\begin{equation}
\log(I(\lambda)) = \log f_u(0) - d \left(\log 4 \sin^2 \frac{\lambda}{2}\right) + \log \frac{f_u(\lambda)}{f_u(0)} + \log \frac{I(\lambda)}{f(\lambda)}
\label{eq:lm_gph}
\end{equation}
where $\log(I(\lambda))$ plays the role of a response variable, $\log f_u(0)$ plays the role of the intercept, and $d$ plays the role of the slope. The term $\log \frac{f_u(\omega)}{f_u(0)}$ is negligible for $\lambda$ close to zero, and $\log \frac{I(\lambda)}{f(\lambda)}$ represents an error term. For short memory processes, these error terms are approximately uncorrelated and homoscedastic, and approximately exponential random variables with $\mathbb{E}\left[\log \frac{I(\lambda)}{f(\lambda)}\right] = - \gamma$ where $\gamma$ is the Euler constant. However, \cite{robinson_logperiog} shows that this is not valid for long memory processes.

\cite{robinson_logperiog} suggests an alternative response variable:
\begin{equation}
y_{j}^{(K)} = \log \sum_{k=1}^{K} I(\lambda_{j+k-K})
\label{eq:robinson}
\end{equation}
for an integer $K > 1$, and $j=\ell+K, \ell+2K, \ldots, m$. Both $\ell$ and $m$ are chosen to tend to infinity with $n$, but more slowly than $n$, and such that $\ell/m \rightarrow 0$. Then the regression model
\begin{equation}
y_j^{K} = c^{K} - d \log \left( 4 \sin^2 \frac{\lambda_j}{2}\right) + u_j^{K}
\label{eq:lm_reg_r}
\end{equation}
for $j=\ell+K, \ell+2K, \ldots, m$ can be proved to have uncorrelated errors with zero mean, under the assumption of differentiability of the spectral density around zero. For ease of notation, from here on we suppress the $K$ apex. \cite{robinson_logperiog} proves asymptotic normality of the least-square estimator of $c$ and $d$ under the assumption of Gaussianity of the initial stochastic process $\{x_t\}_{t \in \mathcal{T}}$. 

\cite{robinson_logperiog} defines the following assumptions to obtain limiting covariance properties of the error terms, in particular, limiting independence.

\textbf{Assumption 1.} There exist $C \in (0,\infty)$, $d \in (-1/2,1/2)$, and $\alpha \in (0,2]$ such that
$$
f(\lambda) = C \cdot \lambda^{-2 d} + O(\lambda^{\alpha-2d}) \qquad \text{ as } \lambda \rightarrow 0^+.
$$
 
\textbf{Assumption 2.} In a neighborhood $(0,\varepsilon)$ of the origin, $f(\lambda)$ is differentiable and
$$
\left| \frac{d}{d\lambda} f(\lambda)\right| = O(\lambda^{-1-2d}) \qquad \text{ as } \lambda \rightarrow 0^+.
$$
No assumptions are imposed on $f$ outside a neighborhood of the origin, apart from integrability implied by covariance stationarity. Assumptions 1 and 2 hold, for example, with $\alpha = 2$ for fractional noise and ARFIMA models.

\begin{figure}[h]
\begin{center}
\includegraphics[width=10cm,height=7cm]{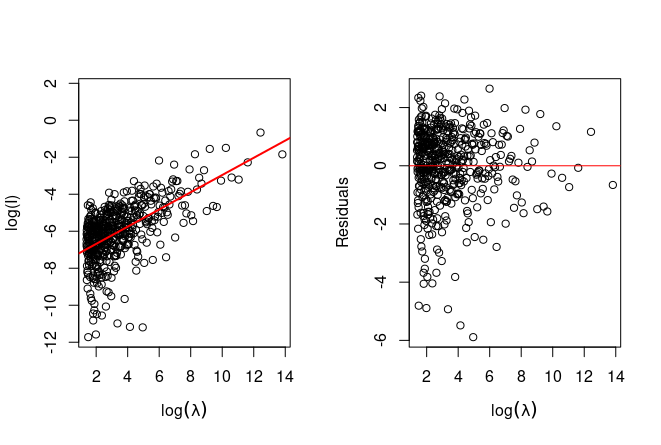}
\end{center}
\caption{Scatterplot of model \eqref{eq:lm_gph} for simulations from an ARFIMA$(0,0.49,0)$ model, and corresponding residuals.}
\label{fig:longmem_reg}
\end{figure}

\begin{figure}[h]
\begin{center}
\includegraphics[width=10cm,height=7cm]{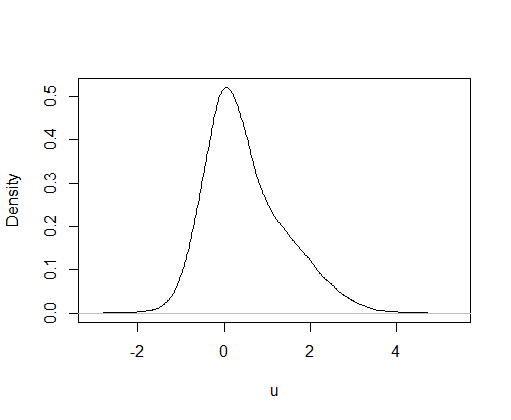}
\end{center}
\caption{Density of the model $p_1(x) N(1,1^2) + p_2(x) N(0,0.5^2)$, where $p_1(x)=1$ if $x>0$ and $p_2(x)=1$ if $x \leq 0$. The covariate has a standard normal distribution.}
\label{fig:asymm_errors}
\end{figure}

Moreover, Figure \ref{fig:longmem_reg} shows that, even if the errors of model \eqref{eq:lm_reg_r} are uncorrelated, they exhibit some form of asymmetry, which made models based on Gaussian distributiond inappropriate. To avoid model misspecification, we propose not defining a parametric model for the error terms $u_j$; thus, model \eqref{eq:lm_reg_r} is a semi-parametric model, with a parametric part (linear regression with coefficient $d$) and a nonparametric part (weak assumption on the model for the errors $u_j$). In particular, asymmetric continuous errors can be represented by an infinite mixture of Gaussian distributions with weights dependent on a covariate, as shown in Figure \ref{fig:asymm_errors}. Therefore, it is possible to represent the model for the log-periodogram via a dependent Dirichlet process mixture (DDPM) model prior on the nonparametric part of the model, with weights defined as functions of the frequencies $\lambda_j$.

From here on, we make use of an approximation $\lambda_j^{-2d} \approx   \left( 4 \sin^2 \frac{\lambda_j}{2}\right)^{-d}$, obtained through a Taylor expansion \citep{hassler2018time} for easy of notation; however, the same results can be obtained using the original values. The data consists of $m$ observations $\{(y_1,\lambda_1),\ldots, (y_m,\lambda_m)\}$. The observations are independent and identically distributed $(y_j, \lambda_j) \sim F_0$ under the assumptions that the data generating process satisfies $\mathbb{E}_{F_0}[y | \lambda] = c - d \log \lambda$ for all $\lambda \in (0,\pi]$. Notice that in practice $\lambda$ assumes values in the Fourier frequencies, therefore $2\pi/n \leq \lambda \leq \pi$, therefore $\log \lambda$ has itself a bounded domain $\Lambda = \{2\pi/n, \ldots, \pi\}$. Alternatively, the restricted moment model can be rewritten as 
$$
y_j = c - d_0 \log \lambda_j + u_j, \qquad (y_j,\lambda_j) \sim F_0, \qquad j=1, \ldots, m
$$
with $\mathbb{E}_{F_0}[u | \lambda] = 0$ for all $\lambda \in \Lambda$. Then the space of conditional densities for $u_j$ is $\mathcal{F}_{u|\lambda}$
$$
\mathcal{F}_{u|\lambda} = \left\{ f_{u | \lambda}: \mathbb{R} \times \Lambda \rightarrow [0,\infty): \int_{\mathbb{R}} f_{u|\lambda}(u,\lambda) du = 1, \int_{\mathbb{R}} u f_{u|\lambda}(u,\lambda) du = 0 \quad \forall \lambda \in \Lambda\right\}.
$$ 

First, we redefine the mixture model in such a way that the mean is assured to be zero. Given a set of parameters ${\pi, \mu, \sigma_1, \sigma_2}$, the density $b$ is defined as
$$
b(u; \pi, \mu, \sigma_1, \sigma_2) = \pi N(u; \mu, \sigma_1^2) + (1-\pi) N\left(u; -\mu \frac{\pi}{1-\pi}, \sigma_2^2\right).
$$
By construction, the random variable $u$ with probability density $b$ has an expected value equal to zero. Then the model for $y$ becomes
\begin{equation}
f(y|\lambda, c, d, \eta) = \sum_{h=1}^{\infty} p_h(\lambda) b(y - c + d \log \lambda; \pi, \mu_h,\sigma_{1h}, \sigma_{2h})
\label{eq:infinitemix}
\end{equation}
such that $\sum_{h=1}^{\infty} p_h(\lambda) = 1$ for all $\lambda \in \Lambda$.
The weights are obtained via a frequency-dependent stick-breaking construction. The kernel stick-breaking process \citep{dunson2008kernel, grazian2024spatio} is defined using a countable sequence of mutually independent random components $V_h \sim Be(a_h, b_h)$ and atoms $\theta_h \sim G$ for each $h = 1, 2, \ldots$, where $Be(a, b)$ stands for a beta distribution and $G$ is a Dirichlet process base distribution. The weights of the mixture are then defined such that, for all $\lambda \in \Lambda$, $p_1(\lambda) = V_1(\lambda)$ and
$$
p_h(\lambda) = V_h(\lambda) \prod_{j=1}^{h-1}(1-V_j(\lambda)) \qquad h=2, 3, \ldots
$$
where $V_h(\lambda) = w_h(\lambda, \psi) V_h$. The atoms of the stick-breaking process are considered not frequency-dependent; therefore, this is an example of a single-atom DDPM.

The mixture weights depend on a bounded kernel function $w: \Lambda \times \Lambda \rightarrow [0,1]$, which gives similar weights to similar frequencies according to some hyper-parameters. For example, the squared exponential kernel can be used:
$$
w_h(\lambda, \psi_h) = \exp\left[ -\frac{(\lambda - \psi_h)^2}{\xi^2}\right],
$$
for some $\xi > 0$ and $\psi_h \in \Lambda$ for all $h$.
The kernel stick-breaking prior can be proved to be proper if $\mathbb{E}[V_h]$ and $\mathbb{E}[w_h(\lambda,\psi)]$ are both positive, which are a assumption easy to obtain in practice, using a beta prior for $V_h$ and bounded kernels. Then Equation \eqref{eq:infinitemix} is well-defined with $\sum_{h=1}^{\infty} p_h(\lambda) = 1$.

The nuisance parameters are given by $\eta = (\xi, \{\pi_h, \mu_h, \sigma_{1h}, \sigma_{2h}, \psi_h\}_{h=1}^{\infty})$, and jointly they induce a prior $\Pi_{\mathcal{H}}$. Specifically, $\xi \sim \text{IG}(a_{\xi}, b_{\xi})$, $\{\mu_h, \sigma_{1h}, \sigma_{2h}\}$ are taken from a normal-inverse gamma prior distribution, $\psi_h \sim \text{Unif}_\Lambda$, $V_h \sim \text{Be}(a_h, b_h)$. Finally, $c \sim N(0,\sigma^2_{c})$ and $d \sim \text{Unif}(-1,1/2)$. This model allows the marginal posterior distribution for $c$ and $d$ to concentrate on the true parameter $(c_0,d_0)$, and on a weak neighborhood of the true conditional densities $f_{0,u|\lambda}$.

\subsection{Consistency properties} 

Consistency results are available in a semiparametric setting using Dirichlet process mixtures (DPM) when errors are independent from the covariates \citep{chib2010additive}, or with errors dependent on covariates but symmetric \citep{pati2014bayesian}. These methods are not appropriate in the presence of heteroscedasticity or asymmetry \citep{pelenis2014bayesian}. We propose another approach inspired by \cite{norets2010approximation}. 

Define $f(y|\lambda)$ as the conditional density of $y$ given $\lambda$. Define the true generating process as $f_0(y|\lambda)$. The Kullback-Leibler (KL) divergence between the true generating model and the proposed model is defined as
$$
d_{KL}(f_0,f) = \int \log \frac{f_0(y|\lambda)f_0(\lambda)}{f(y|\lambda)f(\lambda)} F_0(dy,d\lambda).
$$

We say that the posterior is consistent for estimating $(f_{0,u|\lambda},c_0,d_0)$ if $\Pi(W|y,\lambda) \rightarrow 1$ in $P_{F_0}$-probability as $m \rightarrow \infty$ and for any neighborhood $W$ of $(f_{0,u|\lambda},c_0,d_0)$. Define a weak neighborhood
\begin{footnotesize}
\begin{align*}
\mathcal{U}_{\delta}(f_{u|\lambda}) &= \left\{ f_{|\lambda}: f_{|\lambda} \in \mathcal{F}_{u|\lambda}, \left| \int_{\mathbb{R} \times \Lambda} g f_{|\lambda}(u,\lambda) f_0(\lambda) du d\lambda - \int_{\mathbb{R} \times \Lambda} g f_{u|\lambda}(u, \lambda) f_0(\lambda) du d\lambda \right|< \delta, g: \mathbb{R} \times \Lambda \rightarrow \mathbb{R} \right\}
\end{align*}
\end{footnotesize}
where $g$ is a bounded and uniformly continuous function. $W$ is of the form
$$
W = \mathcal{U}_{\delta}(f_{0,u|z}) \times \{\beta: \parallel \beta-\beta_0\parallel <\rho\}
$$
with $\delta>0$, $\rho>0$, $\beta=(c,d)$ and $\beta_0 = (c_0,d_0)$. 

Consider the following set of assumptions:

\textbf{Assumptions:}
\begin{enumerate}
	\item $\mathbb{E}[y | \lambda] = c - d \log \lambda$ almost surely in $F_0$
	\item $f_0(y|\lambda)$ is continuous in $(y,\lambda)$ almost surely in $F_0$
	\item $b(y;\mu_h, \sigma_h)$ is a fixed symmetric location-scale density with location $\mu_h$ and scale $\sigma_h$, with $\log b(y;\mu_h,\sigma_h)$ integrable with respect to $F_0$
	\item $\mathbb{E}[y^2|\lambda] < \infty$ for all $\lambda \in \Lambda$
	\item there exists $\delta>0$ such that
	$$
	\int \log \frac{f_0(y|\lambda)}{\inf_{\parallel y-s\parallel < \delta,\parallel z-t\parallel < \delta } f_0(s|t)} F_0(dy,d\lambda) < \infty
	$$
\end{enumerate}

Priors $\Pi_{\mathcal{H}}, \Pi_{c}, \Pi_{d}$ and kernel $w_{h}(\lambda,\psi_h)$ induce a model
$$
f(y|\lambda,c,d,\eta) = \sum_{h=1}^{\infty} p_h(\lambda) b(y- c + d \log \lambda; \pi_h, \mu_h, \sigma_{1h},\sigma_{2h})
$$
with weights $p_1 = V_1$ and $p_h(\lambda) = V_h w(\lambda,\psi_h) \prod_{\ell < h}(1- V_{\ell} w(\lambda, \psi_{\ell}))$ for $h\geq 2$.  

\begin{theorem}
Assume $F_0$ satisfies assumptions 1-4, and $f_0(\cdot | \cdot)$ are covariate-dependent conditional densities of $y \in \mathcal{Y}$ induced by $F_0$. Let $w(\lambda,\psi)$ be the squared exponential kernel and let $\Pi_{\mathcal{H}}$ be induced by the kernel stick-breaking process. If priors are such that $c_0$ and $d_0$ are a interior point of the support of $\Pi_{c}$ and $\Pi_{d}$ respectively, and $\Pi(\sigma_{1h}<\delta)>0$ for any $\delta>0$, then $f_0 \in KL(\Pi)$, i.e.
$$
\Pi\left( \{ (c, d,\eta): d_{KL}(f_0(\cdot | \cdot), f(\cdot | \cdot, c, d, \eta))< \varepsilon \}\right)>0
$$ 
where $f(y | \lambda, c, d, \eta) = \sum_{h=1}^{\infty} p_h(\lambda) b(y - c + d \log \lambda; \pi_h, \mu_h, \sigma_{1h}, \sigma_{2h})$, $p_1 = V_1$ and $p_h(\lambda) = V_h w(\lambda, \psi_h) \prod_{\ell < h} (1-V_{\ell} w(\lambda, \psi_{\ell}))$ for $h\geq 2$.
\label{th:approx}
\end{theorem}

This indicates that there exists a set of nuisance parameters within the infinite model that approximate the final model in terms of Kullback-Leibler (KL) divergence. In other words, any parameter set within a neighborhood of this specific instance of the infinite model is arbitrarily close to it in terms of KL divergence, when this neighborhood has positive prior support.

\begin{theorem}
Suppose $F_0$ satisfies assumptions 1-4 and that $Var(\lambda)>0$. The prior is constrained so that there exists a large $L$ such that $\mathbb{E}f[u^2 |\lambda] < L$ for any $\lambda \in \Lambda$ and all $f \in supp(\Pi_{f_{u|\lambda}})$. Define
$$
W = \mathcal{U}_{\delta}(f_{0,u | \lambda}) \times \{\beta: \parallel \beta-\beta_0\parallel<\rho\}
$$
with $\delta>0$, $\rho>0$, $\beta=(c,d)$ and $\beta_0=(c_0,d_0)$. 
Then $\Pi(W^C | y,\lambda) \rightarrow 0$ almost surely in $P_{F_0}$-probability as $m \rightarrow \infty$.
\label{th:consistency}
\end{theorem}

This indicates that as the sample size increases, both the conditional distribution of the error terms and the Bayesian estimator of $c$ and $d$ converge to the true distribution and values.

An important fact of the results in this Section is that no assumption on the distribution of the stochastic process has been done, while other works have often used the assumption of Gaussianity to obtain asymptotic results. 

Proofs of Theorem \ref{th:approx} and \ref{th:consistency} are provided in the Appendix.

\section{Simulation study}
\label{sec:simu}

In order to assess the performance of the proposed modeling approach, we designed five different scenarios with increasing values of the long memory parameter, resulting in a total of 50 scenarios. For each scenario, we consider 50 replicates, meaning that multiple time series are generated from the same underlying process.

The approach proposed in this work successfully estimated the parameter of long memory. For each time series, we simulated $n=10,000$ time points, resulting in 5,000 observations from the log-periodogram. We investigated posterior estimates based on the posterior mean, for $d$ and credible intervals, as well as coverage.

We first consider a simple autoregressive fractionally integrated model, $\text{ARFIMA}(0,d,0)$:
$$
(1-B)^d X_t = \varepsilon_t
$$ 
where $B$ is the backward shift operator; the relative spectral density is
$$
f(\lambda) = \frac{1}{2\pi} \left(2 \sin\left(\frac{\lambda}{2}\right)\right)^{-2d}.
$$
The error terms $\varepsilon_t$ are independent and identically distributed from a normal distribution with mean zero and variance equal to one.

The other four scenarios consider $\text{ARFIMA}(p,d,q)$ models with $p=1$ and $q=1$:
$$
\left( 1 - \phi B \right)(1-B)^d X_t = (1 + \theta B) \varepsilon_t,
$$
where $\phi$ is the parameter of the autoregressive part of the model, and $\theta$ is the parameter of the moving average part of the model.
The corresponding spectral density is then:
\begin{align*}
f(\lambda) &= \frac{\sigma^2}{2\pi} \frac{|\theta(e^{-i\lambda})|^2}{|\phi(e^{-i\lambda})|^2} |1-e^{-i\lambda}|^{-2d}.
\end{align*}
Again, the error terms are generated i.i.d. from a normal distribution with mean zero and variance one. The choice of the AR parameter $\phi$ and the MA parameter $\theta$ is taken to consider cases with increasing importance of the autoregressive or moving average part: $(\phi=0.8, \theta=0.1)$, $(\phi=0.5, \theta=0.1)$, $(\phi=0.1, \theta=0.5)$, and $(\phi=0.1, \theta=0.8)$.

We assumed a truncated approximation of the Dirichlet process with the number of mixture components fixed to $H=30$; similar results were obtained with both larger and smaller values of $H$. We assume $c$ to have a normal prior distribution with mean zero and variance $1,000$, and $d$ to have a uniform prior distribution in $(-1,1/2)$. A stick-breaking prior process is considered, with $a_h=a$ and $b_h=b$ for $\forall h$, where $a \sim Unif(0,10)$ and $b \sim Unif(0,10)$. The base distribution of the Dirichlet process is assumed to be a normal distribution with mean zero and variance $ \sigma^2_{\theta}\sim IG(0.01,0.01)$. The knots $\psi_h$ are assumed to follow a uniform prior distribution in $\Lambda$. A double exponential kernel is used for $w_h(\lambda,\psi_h)$, with $\xi \sim IG \left( 1.5, \frac{\nu^2}{2}\right)$ and $\nu \sim Unif(0,1)$.

The proposed approach is compared with the least squares estimator defined by \cite{robinson_logperiog} using the \texttt{gph()} function of the \texttt{LongMemoryTS} R package \citep{leschinski2019longmemoryts}, fixing $\ell=1$ (following the default choice of the \texttt{gph()} function) and $m=1+n^{0.8} = 1,586$. 

Moreover, the proposed approach was compared to a parametric Bayesian approach, where the correct $\text{ARFIMA}(0,d,0)$  or $\text{ARFIMA}(1,d,1)$ model were assumed and Bayesian inference was run for the parameters $(\phi,d,\theta)$ via a Monte Carlo Markov Chain (MCMC) approach, choosing $\sigma^2 \sim IG(0.1,0.1)$, and considering uniform priors for $\phi$ and $\theta$ under the constraint that $x_t$ is stationary and $x_t$ is invertible, respectively. We follow the approach of \cite{palma2007long} and \cite{durham2019bayesian}. 

\begin{figure}
\centering
\includegraphics[width=13cm,height=7cm]{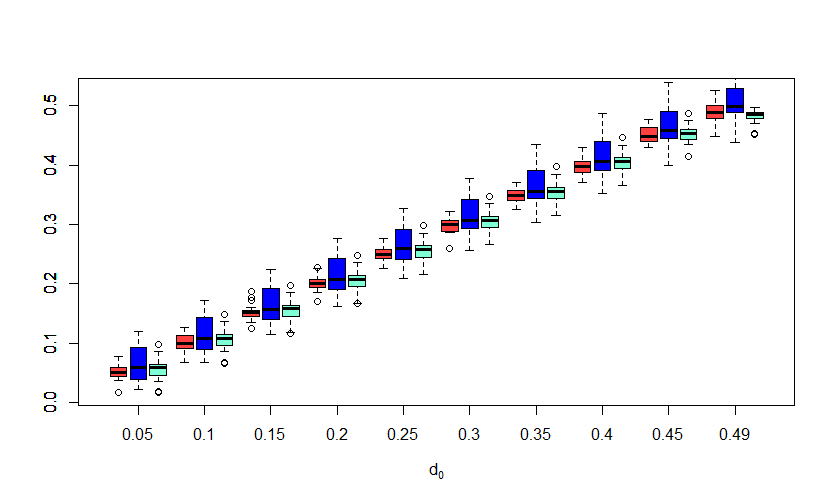}
\caption{Boxplotes of estimation of $d$ for different methods (red is for the Bayesian semiparametric approach proposed in this work, blue is for the frequentist estimator of \cite{robinson_logperiog}, and acquamarine is for a parametric approach using the correct ARFIMA model). Simulations are from an ARFIMA$(0,d,0)$ model, with increasing $d$, and boxplots are obtained over 50 repetitions of the experiment for each model.}
\label{fig:boxplots_0d0}
\end{figure}

\begin{figure}
\centering

\begin{minipage}{.50\linewidth}
\centering
\subfloat[]{\label{fig:boxARhighMAlow}\includegraphics[width=8.5cm,height=7cm]{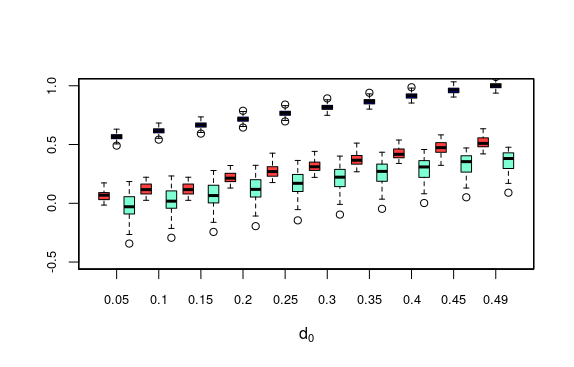}}
\end{minipage}%
\begin{minipage}{0.50\linewidth}
\centering
\subfloat[]{\label{fig:boxARmediumMAlow}\includegraphics[width=8.5cm,height=7cm]{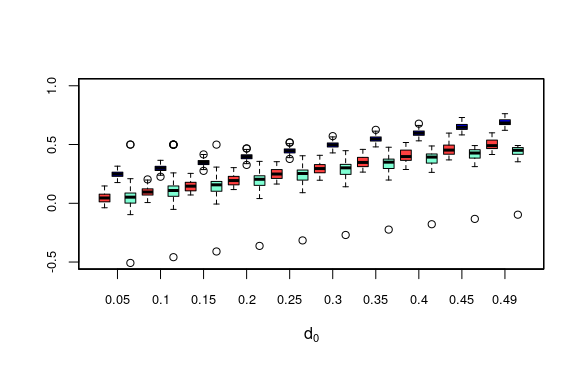}}
\end{minipage}\par\medskip

\begin{minipage}{.50\linewidth}
\centering
\subfloat[]{\label{fig:boxARlowMAmedium}\includegraphics[width=8.5cm,height=7cm]{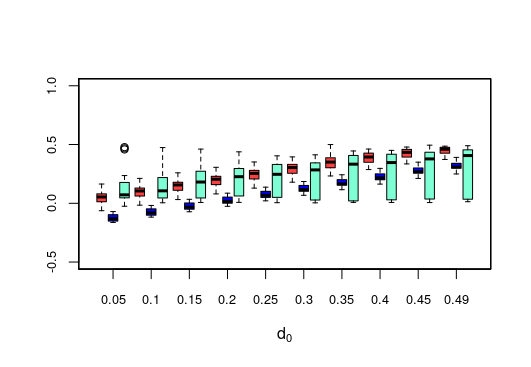}}
\end{minipage}%
\begin{minipage}{0.50\linewidth}
\centering
\subfloat[]{\label{fig:boxARlowMAhigh}\includegraphics[width=8.5cm,height=7cm]{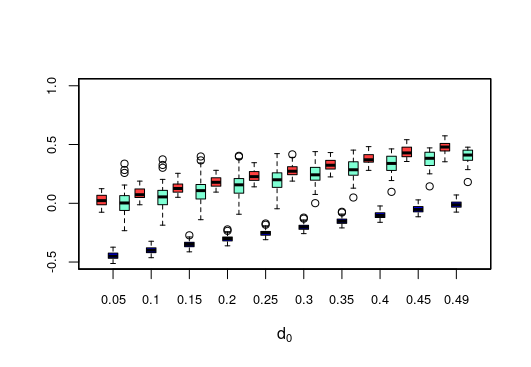}}
\end{minipage}\par\medskip

\caption{Boxplots are constructed as in Figure \ref{fig:boxplots_0d0}, but for ARFIMA$(1,d,1)$ models. (a) refers to a model with $\phi=0.8$ and $\theta=0.1$, (b) refers to a model with $\phi=0.5$ and $\theta=0.1$, (c) refers to a model with $\phi=0.1$ and $\theta=0.5$, and (d) refers to a model with $\phi=0.5$ and $\theta=0.8$}
\label{fig:boxplots_1d1}
\end{figure}

The average estimates and the 95\% intervals over 50 replications are shown in Figures \ref{fig:boxplots_0d0} and \ref{fig:boxplots_1d1}, and in Tables \ref{tab:simu_mine}, \ref{tab:simu_freq}, and \ref{tab:simu_param}. The coverage based on 95\% credible or confidence intervals is shown in Table \ref{tab:simu_coverage}. The proposed semiparametric approach seems to be able to recover the correct value of the parameter of long memory for all scenarios, independently from the presence of short memory. The only case that seems to be more problematic is the case of a low value of long memory ($d=0.05$) and presence of an intermediate moving average component ($\phi=0.1$, $\theta=0.5$). For all other cases, the true value is included in the corresponding intervals obtained over repetitions. The case of an intermediate moving average component is also the case where there is more uncertainty about the parameter of long memory.

The comparison with the least square estimator of \cite{geweke1983estimation} and \cite{robinson_logperiog} in Table \ref{tab:simu_freq} seems to suggest that there is more uncertainty around the value, and the least square estimator seems to have difficulties in identifying the value of the parameter of long memory in the presence of short memory. This may be an indication that, while \cite{robinson_logperiog} proved asymptotic normality and consistency of the least-squares estimator, consistency is reached for larger sample sizes than the ones needed with the proposed Bayesian approach. This behavior seems to worsen in the presence of a large moving average component.

Finally, Table \ref{tab:simu_param} shows the results when assuming a correct parametric model; differently from the frequentist semiparametric estimates, this estimation process seems to show high variability, especially in the presence of short memory, with outliers. Also, this method seems to exhibit the lowest coverage levels, particularly in the presence of short memory. In particular, the chains obtained through MCMC seem to display large autocorrelations, with low effective sample sizes, and the estimation process of $d$ seems to be more affected by the values of the $ARMA$ part. These results are obtained when assuming the correct model; however, the behavior is expected to worsen in the presence of misspecification.

\section{Real data applications}
\label{sec:data}

\subsection{Nile River Data}

A benchmark example of a dataset exhibiting both long memory and possible inhomogeneity in variance is a time series regarding the annual minimum water levels of the Nile River \citep{toussoun1926memoire}. This time series comprises 663 yearly values recorded from 622 AD to 1284 AD at the Roda Gauge near Cairo. Figure \ref{main:nile_ts} displays the relative time series, sourced from the \texttt{longmemo} R package \citep{maechler2020package}. From Figure \ref{main:nile_ts} it is evident that there are extended periods during which the observations tend to remain at either high or low levels. However, there does not appear to be a persistent cycle. Figure \ref{main:nile_acf} illustrates the autocorrelations of the Nile data up to a horizon of 40 lags: the autocorrelations seem to decay at a slow pace.

In his study, \cite{beran_longmem} discovered that a stationary Gaussian fractionally differenced model fits this time series effectively. \cite{beran_longmem} estimated the long memory parameter to be $\hat{d}=0.40$ using an approximate maximum likelihood approach.

The least square estimator provided by \cite{robinson_logperiog} yields a point estimate of $\hat{d}=0.386$. This estimate was computed using the \texttt{gph()} function from the \texttt{longmemo} R package \citep{maechler2020package} with $m=1+n^{0.8} = 252$, where $n$ represents the length of the series. The corresponding 95\% confidence interval is $[0.365,0.590]$.

The proposed Bayesian approach, utilizing the same prior distributions as outlined in Section \ref{sec:simu}, resulted in a posterior mean estimator of $\mathbb{E}(d | x_1, \ldots,x_n) = 0.351$, with credible interval $[0.301,0.421]$. Comparatively, while both the frequentist and the Bayesian nonparametric approaches seem to exclude the hypothesis that $d=0$, the frequentist method does not reject the hypothesis that $d>0.5$. However, the Bayesian approach clearly suggests that $d<0.5$, indicating that the series is stationary and possesses long memory. This inference is further supported by the Whittle likelihood estimator computed by \cite{baum2020local} as $\hat{d}_{WL}=0.409$, with confidence intervals of $[0.347,0.471]$.

\subsection{Financial data}

The presence of long memory in U.S. indexes, such as the S\&P500, suggests that past information significantly influences future returns over extended periods, challenging the efficient market hypothesis. This persistence in index fluctuations implies that historical data, both distant and recent, hold valuable insights for predicting future movements. Factors such as investor behavior, macroeconomic trends, and structural changes in financial markets contribute to this phenomenon. Understanding and modeling long memory effects are crucial for investors and policymakers navigating the complexities of the stock market. Research demonstrates mixed evidence of long memory in the S\&P500 and Nasdaq indexes during such periods, raising important implications for equity return modeling.

For instance, R/S analysis \citep{peters1989fractal} finds evidence for such persistence, but this finding disagrees with the modified version of R/S developed in \cite{lo1991long} and \cite{ambrose1993fractal}. \cite{hays2010evidence} seem to suggest that long memory characterizes periods of large deviations from rational pricing.

In this section, we analyze daily S\&P500 prices between April 22, 2019, and April 19, 2024. Figure \ref{main:sp500_ts} shows the time series for the prices, and Figure \ref{main:sp500_lr} shows the log-returns. Figure \ref{main:sp500a} presents the periodogram, which shows a dominant peak for frequencies above 0.4, corresponding to a period of about 2-3 days. Using a regression model between the log-periodogram and the log-frequencies shows that there seems to be a negative dependence, and the points have an asymmetric distribution around the regression line (Figures \ref{main:sp500b} and \ref{main:sp500c}).

The Hurst exponent for the R/S analysis computed on this dataset provides a value of $H=0.521$, while the corrected version yields $H=0.550$. Additionally, the empirical Hurst exponent is $H=0.550$. All these values seem to suggest there is no long memory, since $H=0.5$ corresponds to a Brownian motion.

The least squares nonparametric estimator of \cite{robinson_logperiog} provides an estimate of $\hat{d}=0.030$ for $m=1+n^{0.8} = 302$, with a 95\% confidence interval of $[-0.221,0.050]$. Therefore, the frequentist analysis does not exclude that $d=0$, suggesting the possibility of antipersistence.

The proposed approach has been applied with the same prior distributions as described in Section \ref{sec:simu}. The obtained posterior estimate (computed as the posterior mean) is $\hat{d}=-0.020$, with credible intervals of $[-0.053,0.106]$. This result seems to agree with the frequentist analysis, where there is not enough evidence to reject $d=0$.

These findings suggest that there is no evidence of long memory in the daily returns series for the S\&P500 index between 2019 and 2024. These results align with the findings of \cite{hays2010evidence}. However, \cite{hays2010evidence} suggest that long memory in financial indices can be exhibited in specific and limited periods, such as the 1990s bubble. Additionally, more recently, \cite{wu2022long} showed that Bitcoin price time series seem to exhibit long memory during the COVID-19 pandemic.

To further investigate the effect of the COVID-19 pandemic on the S\&P500 index, we repeated the analysis for a series limited between July 30, 2019, and October 31, 2020.

With this subset of the initial dataset, while the Hurst exponent remains $H=0.532$, the corrected version increases to $H=0.592$, and the empirical Hurst exponent is $H=0.802$, suggesting that the series exhibits long memory. The frequentist estimate of the long memory parameter provides $\hat{d}=0.114$, with a 95\% confidence interval of $[-0.401,0.253]$.

Applying the proposed Bayesian semiparametric approach, the posterior mean is estimated as $\hat{d}=0.143$, with 95\% credible intervals of $[0.079,0.192]$. These results suggest that the S\&P500 index exhibits low levels of long memory during stressful periods for markets, such as during the COVID-19 pandemic.

It is worth noting that the credible intervals are significantly shorter than the corresponding 95\% confidence intervals of the least square estimator. This allows us to conclude that the reduced time series show low-to-moderate long memory.

\subsection{Global Temperature Data}

The climate system displays significant inertia, with over 90\% of the heat generated by human activities being absorbed by the ocean \citep{von2016imperative}. This delayed response implies that even if emissions were halted immediately, temperatures would continue to rise for decades. Understanding and predicting this reaction is critical for devising effective adaptation strategies.

Research \citep{koscielny1998indication, dangendorf2014evidence, franzke2020structure} has identified long-term climate memory, suggesting that climate variables exhibit persistence over extended periods. Various methods, such as Fluctuation Analysis \citep{peng1992long} and Multibox Energy Balance Models \citep{fredriksen2017long}, have been developed to quantify this phenomenon. However, uncertainty still remains regarding the level of long memory.

The data we will analyze in this section are the HadCRUT5.0.1 monthly global mean surface temperature anomalies, publicly available from the Met Office Hadley Centre at \url{https://www.metoffice.gov.uk/hadobs/hadcrut5/data/HadCRUT.5.0.2.0/download.html}. The dataset comprises 2,090 monthly observations spanning from January 1850 to February 2024.

\begin{figure}
\centering

\begin{minipage}{.30\linewidth}
\centering
\subfloat{\label{main:empty}\includegraphics[width=5cm,height=5cm]{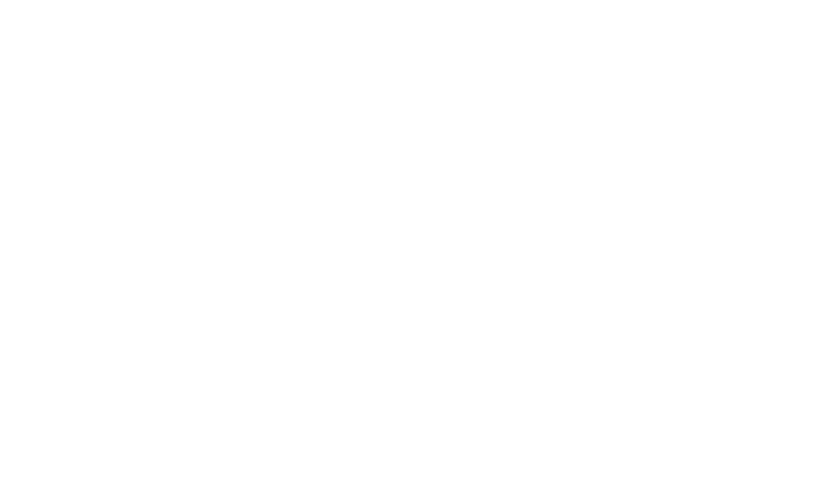}}
\end{minipage}%
\begin{minipage}{0.30\linewidth}
\centering
\subfloat[]{\label{main:nile_ts}\includegraphics[width=5cm,height=5cm]{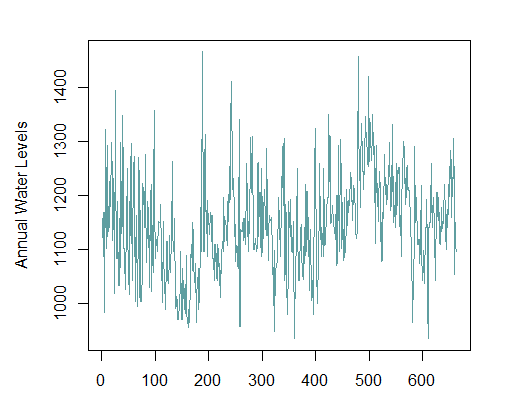}}
\end{minipage}%
\begin{minipage}{.30\linewidth}
\centering
\subfloat[]{\label{main:nile_acf}\includegraphics[width=5cm,height=5cm]{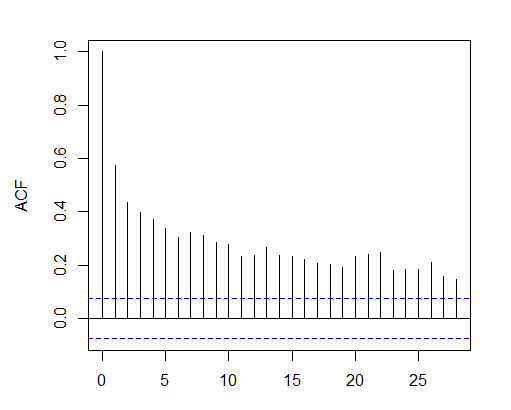}}
\end{minipage}\par\medskip

\begin{minipage}{0.30\linewidth}
\centering
\subfloat[]{\label{main:sp500_ts}\includegraphics[width=5cm,height=5cm]{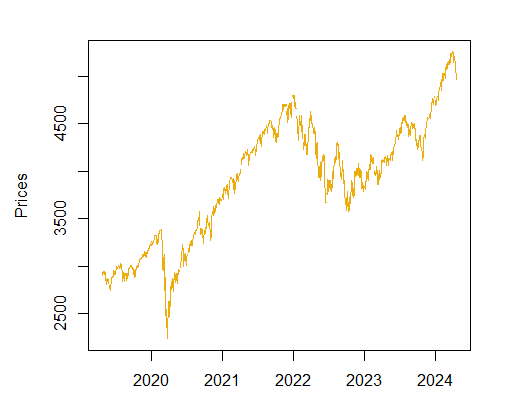}}
\end{minipage}%
\begin{minipage}{.30\linewidth}
\centering
\subfloat[]{\label{main:sp500_lr}\includegraphics[width=5cm,height=5cm]{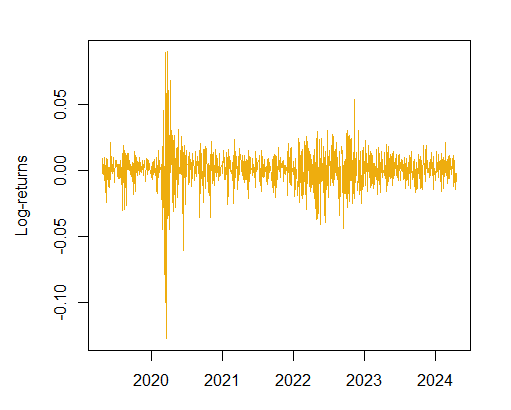}}
\end{minipage}%
\begin{minipage}{.30\linewidth}
\centering
\subfloat[]{\label{main:sp500_acf}\includegraphics[width=5cm,height=5cm]{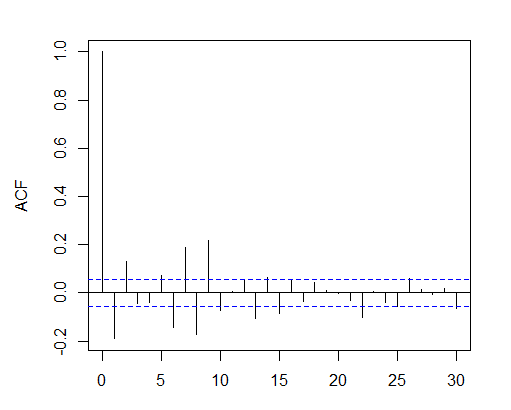}}
\end{minipage}\par\medskip

\begin{minipage}{0.30\linewidth}
\centering
\subfloat[]{\label{main:temp_ts}\includegraphics[width=5cm,height=5cm]{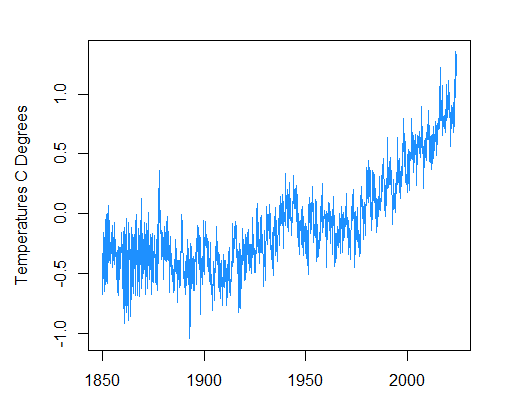}}
\end{minipage}%
\begin{minipage}{.30\linewidth}
\centering
\subfloat[]{\label{main:temp_diff}\includegraphics[width=5cm,height=5cm]{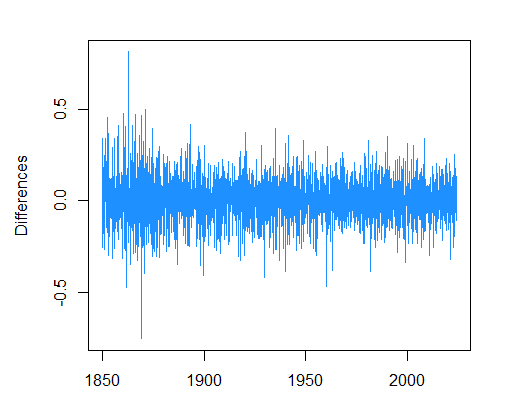}}
\end{minipage}%
\begin{minipage}{.30\linewidth}
\centering
\subfloat[]{\label{main:temp_acf}\includegraphics[width=5cm,height=5cm]{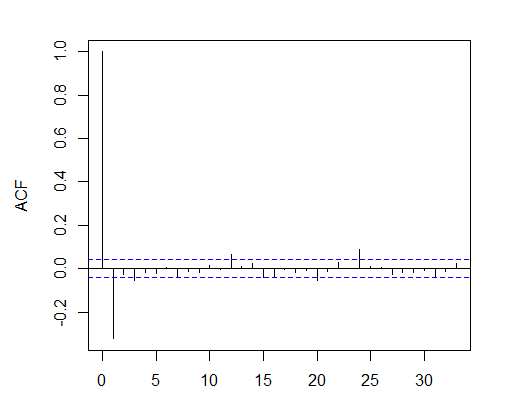}}
\end{minipage}\par\medskip

\caption{Original time series (left) and the time series used in the analysis (right) for the three datasets considered in this work: Nile dataset (top), S\&P500 dataset (middle) and global temperatures dataset (bottom).}
\label{fig:timeseries}
\end{figure}

Figure \ref{main:temp_ts} displays the observed series from January 1850 to February 2024. There is a clear rise in the average temperature over the last two centuries, with data before 1850 exhibiting higher variability likely due to differences in temperature registration methods and fewer available measurements, which will not be considered in the analysis. In the last 10–20 years, it seems that the increase in average monthly temperatures has proceeded at a steeper rate.

Figure \ref{main:temp_diff} illustrates the differenced series: while the original time series shows long-standing autocorrelation, the differenced series exhibits much lower autocorrelation.

\begin{figure}
\centering

\begin{minipage}{0.30\linewidth}
\centering
\subfloat[]{\label{main:nilea}\includegraphics[width=5cm,height=5cm]{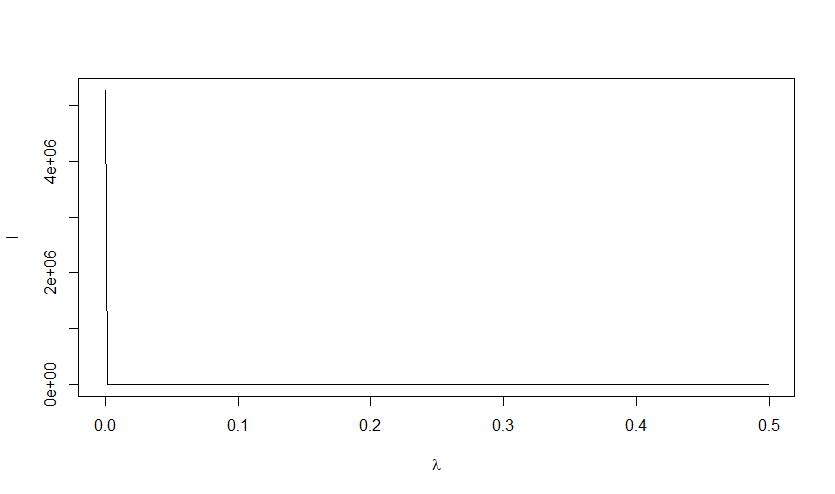}}
\end{minipage}%
\begin{minipage}{.30\linewidth}
\centering
\subfloat[]{\label{main:nileb}\includegraphics[width=5cm,height=5cm]{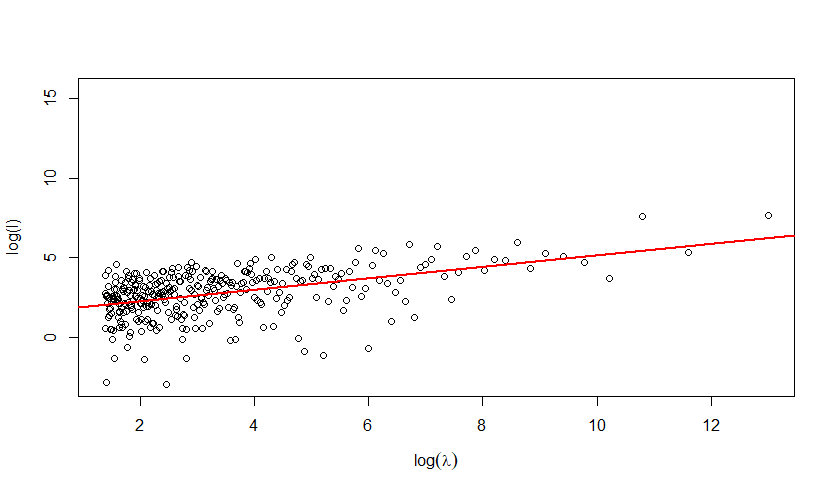}}
\end{minipage}
\begin{minipage}{.30\linewidth}
\centering
\subfloat[]{\label{main:nilec}\includegraphics[width=5cm,height=5cm]{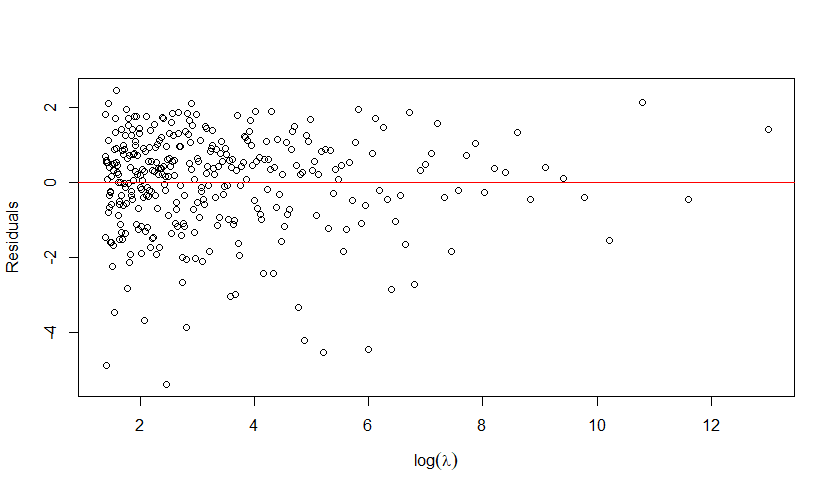}}
\end{minipage}\par\medskip

\begin{minipage}{0.30\linewidth}
\centering
\subfloat[]{\label{main:sp500a}\includegraphics[width=5cm,height=5cm]{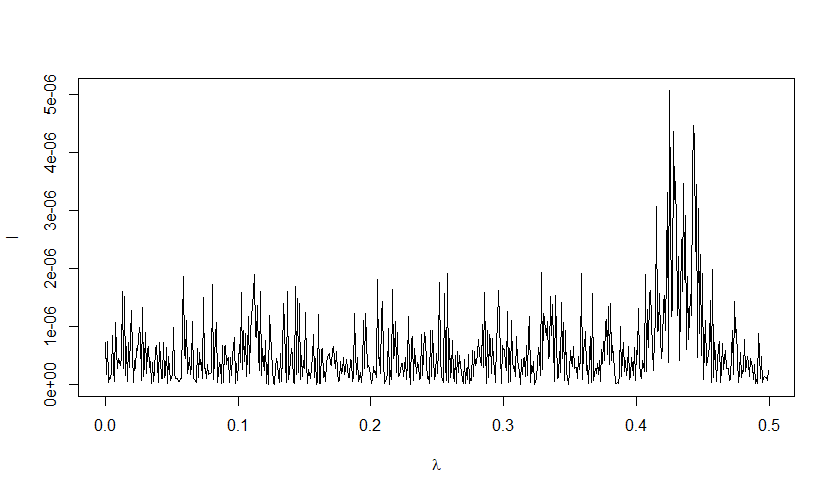}}
\end{minipage}%
\begin{minipage}{.30\linewidth}
\centering
\subfloat[]{\label{main:sp500b}\includegraphics[width=5cm,height=5cm]{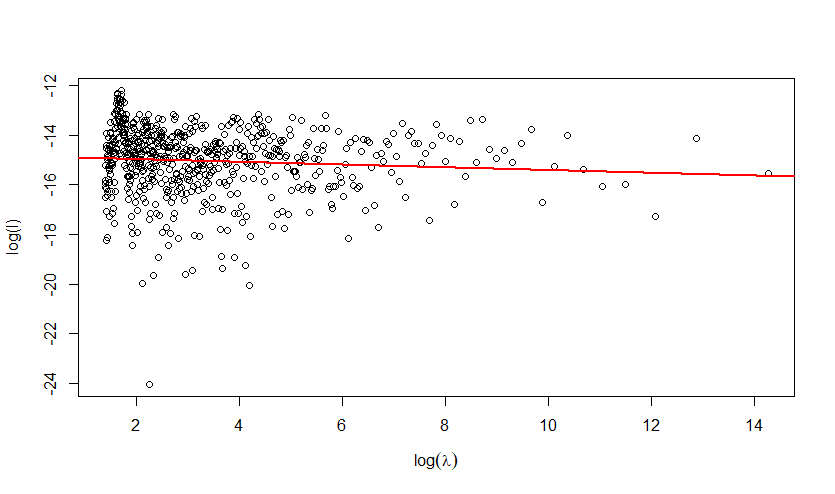}}
\end{minipage}
\begin{minipage}{.30\linewidth}
\centering
\subfloat[]{\label{main:sp500c}\includegraphics[width=5cm,height=5cm]{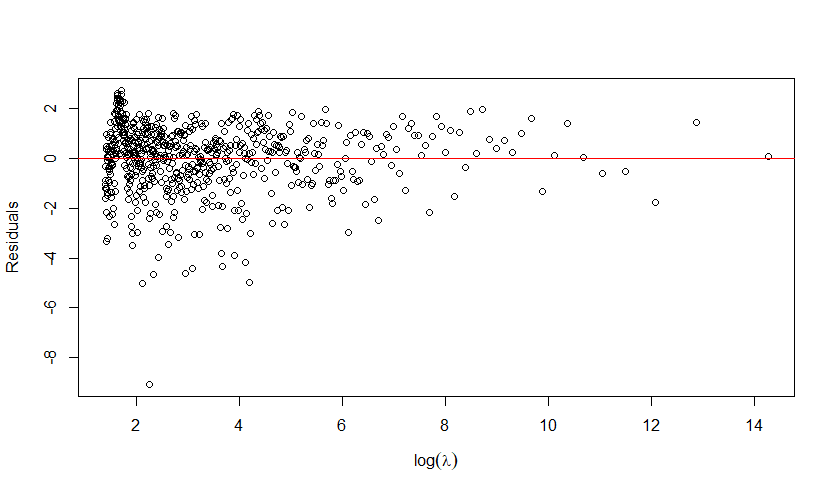}}
\end{minipage}\par\medskip

\begin{minipage}{0.30\linewidth}
\centering
\subfloat[]{\label{main:tempa}\includegraphics[width=5cm,height=5cm]{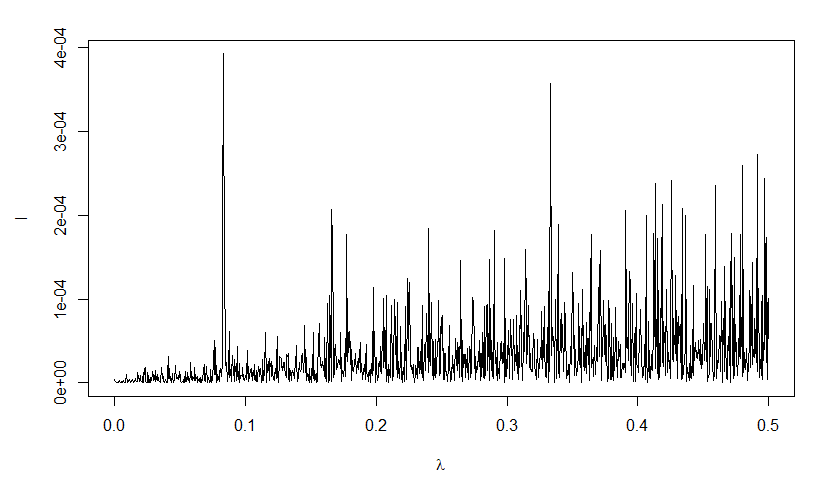}}
\end{minipage}%
\begin{minipage}{.30\linewidth}
\centering
\subfloat[]{\label{main:tempb}\includegraphics[width=5cm,height=5cm]{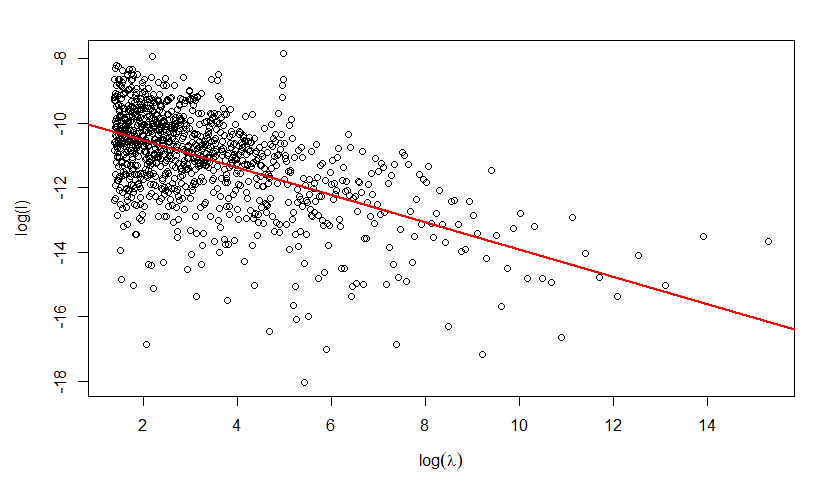}}
\end{minipage}
\begin{minipage}{.30\linewidth}
\centering
\subfloat[]{\label{main:tempc}\includegraphics[width=5cm,height=5cm]{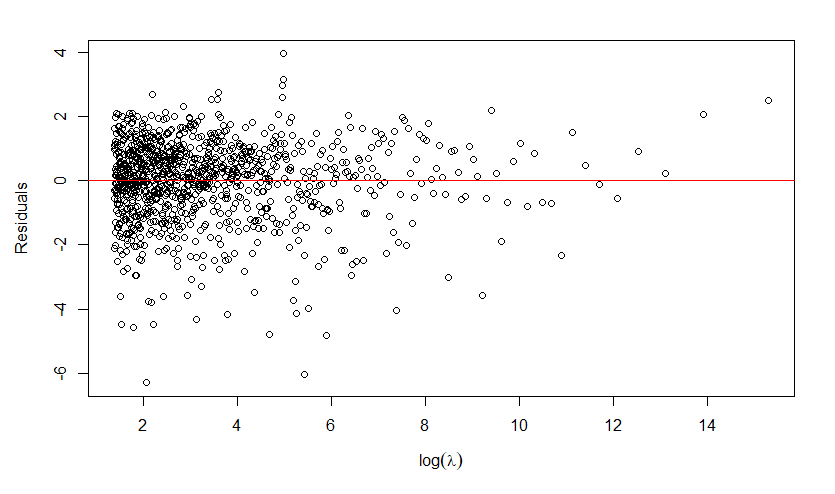}}
\end{minipage}\par\medskip

\caption{Periodograms (left), log-periodogram scatterplost and regression lines with respect to the log-frequencies (middle), and residual plots of the regression model (right) for the three datasets considered in this work: Nile dataset (top), S\&P500 dataset (middle) and global temperatures dataset (bottom).}
\label{fig:periodregres}
\end{figure}

The periodogram for the monthly series (Figure \ref{main:tempa}) shows a peak at $\lambda=0.1$, with the dominant peak area occurring somewhere between frequencies of 0.3 and 0.5. This corresponds to periods of about $1/0.1=10$ years and between 2 and 3.5 years.

The log-periodogram values with respect to the log-frequencies are shown in Figure \ref{main:tempb}: there is a clear negative linear relationship, and the residuals (Figure \ref{main:tempc}) show asymmetric errors, as we expect from Section \ref{sec:meth}. 

Following \cite{yuan2022impact}, we implemented a Detrended Fluctuation Analysis of second order (DFA2) on the monthly global temperatures. DFA2 calculates the fluctuation function $F(s)$: if it increases with the time scale $s$ as a power law, $F(s) \sim s\alpha$, with $\alpha > 0.5$, then long-term climate memory is present in the time series. In this case, a clear power law relation between $F(s)$ and $s$ is found with a slope of 0.90, which may be indicative of the presence of long memory. However, \cite{bashan2008comparison} and \cite{holl2016detrended} show that DFA can obtain similar results for both long memory series and series with a drift.

The Hurst exponent for the R/S analysis provides a value $H=0.332$, the corrected version $H=0.288$, and the empirical Hurst exponent is $H=0.221$. These values are indicative of an anti-persistent behavior, i.e., a tendency for a time series to revert to its long-term mean value.

The least square estimator gives a point estimate of $\hat{d}=-0.434$, with $m=1+n^{0.8} = 453$, where $n$ is the length of the series, and a 95\% confidence interval $[-0.437,0.063]$. These results seem to suggest that the process has short memory and it is possible that the process is anti-persistent at the same time, however the value $d=0$ is not rejected. The proposed approach has been applied to this dataset and the estimated posterior mean resulted in $\mathbb{E}[d|x_1, \ldots,x_n]=-0.351$, with 95\% credible intervals $[-0.415,-0.295]$, again supporting more strongly the idea of anti-persistence in the differenced series.

Similarly to \cite{yuan2022impact}, these results suggest that the process of monthly global temperatures is influenced by what \cite{yuan2022impact} call the ``accumulated responses'' (or indirect memory-response), which is due to external factors, while the direct memory (which is the internal process long memory) does not seem to have an effect on the memory of the process.

\section{Conclusion}
\label{sec:conc}

We have developed a Bayesian semiparametric methodology to estimate the parameter of long memory in a univariate stochastic process with unknown distribution. This problem has received considerable attention in the frequentist literature but limited attention across Bayesian approaches, particularly without the assumption of Gaussianity of the processes.

While other semiparametric Bayesian approaches exist, they often come at a high computational cost or require assumptions about the particular stochastic process. Our approach is based on the definition of the regression model leading to the least squares estimator of \cite{robinson_logperiog}, but it represents the unknown distribution of the error terms through a mixture of Gaussian distributions with frequency-dependent weights. This results in a flexible mixture model capable of representing the typical asymmetry of the error terms. Such asymmetry is sometimes represented via Gamma distributions, but this assumption is valid only for short memory processes. In this work, we propose to represent the unknown distribution of the errors via a single-atom dependent Dirichlet process in the definition of the kernel stick-breaking process, using exponential kernels to define the weights. This is a fundamental feature to obtain consistency results. An additional advantage is that the mixture model framework allows for straightforward implementation of an MCMC algorithm.

The proposed approach is flexible because it limits the assumptions on the stochastic model and allows for the consideration of cases where $d \in (-1,1/2)$, thereby considering cases of long memory and anti-persistence. Through simulation studies, we have demonstrated the model's ability to recover the correct value of the long-memory parameter in the presence of short memory or not. We also applied the methodology to three different datasets, obtaining results that are in agreement with similar field research.

Methods based on the Whittle likelihood are asymptotically correct for both Gaussian and non-Gaussian time series \cite{hannan1973asymptotic}, but they involve an assumption of asymptotic independence of the Fourier coefficients, which is not valid in the case of long-memory processes. Recent works \citep{kirch2019beyond,sykulski2016biased} propose nonparametric corrections to address this issue. Our methodology does not rely on the use of the log-periodogram but on a modified log-periodogram, relating its coefficients to the frequencies of the spectral density through a linear regression model where the error terms are assumed to be asymptotically independent. Differently from many works, including \cite{robinson_logperiog} in a frequentist setting and \cite{clara_bayes} in a Bayesian setting, no assumption of Gaussianity of the process is required to obtain consistency result of the estimator of $d$. 

With respect to \cite{liseo2001bayesian}, the spectral density for the short-memory part is modeled on the frequency band $(0,\pi]$, while in their work, it is defined on $[\lambda^*,\pi]$, which introduces the additional problem of choosing $\lambda^*$. Compared to the model of \cite{Holan2009ABA}, we do not require the assumption of Gaussianity of the process. Additionally, \cite{Holan2009ABA} includes the mean of the process, which is an important characteristic since \cite{hurvich2002multistep} show that estimating the mean is usually slow for $d \in (0,1/2)$. On the other hand, our approach has the advantage of being performed in the frequency domain, with the periodogram being invariant to additive constants.

In this work, we have established the consistency of the Bayesian estimator of $d$ with respect to $d_0$, the true value of long memory parameter, as well as the consistency of the conditional distribution of the errors given the frequencies. Several authors have studied the asymptotic properties of Bayesian semiparametric models. \cite{bickel2012semiparametric} prove a Bernstein-von Mises theorem for semiparametric models under several assumptions, including asymptotic equicontinuity, asymptotic tightness, and smoothness of the mode. \cite{panov2015finite} investigate the finite-sample properties. However, the assumptions required to ensure a Bernstein-von Mises result are challenging to verify in practice for a wide range of models. \cite{castillo2012gaussian}, \cite{castillo2012semiparametric}, and \cite{castillo2015bernstein} study Gaussian process priors, \cite{rivoirard2012bernstein} focus on linear functionals of densities, while \cite{pati2014bayesian}, \cite{chae2019bayesian} and \cite{chae2019semi} demonstrate consistency in the presence of symmetric errors.

A central characteristic of the regression model used in this work is its asymmetric errors. Therefore, these existing results are not easily applicable, and we leave addressing this issue as a topic for future development.

Extensions to multivariate processes are left for further research as well. While theory on the error terms of the regression model used in this approach exists \citep{robinson_logperiog} and requires just a few additional assumptions, deriving the posterior distribution and studying its asymptotic properties necessitates an investigation of multivariate dependence, which is a non-trivial extension of the proposed approach.

Interesting future directions include extending the methodology to nonstationary time series. Simulations not shown in this work suggest that the true values of $d$ are recovered even when $d>1/2$; however, classical results on long-memory processes are typically based on the assumption of weak stationarity. Future research will focus on expanding the mixture model in such a way that nonstationarity can be accommodated.

\bigskip
\begin{center}
{\large\bf SUPPLEMENTARY MATERIAL}
\end{center}

\begin{description}

\item[Title:] Proof of Theorem \ref{th:approx}. 

Let $\varepsilon>0$. Consider a finite mixture model 
$$
f_{fin}(y | \lambda, d, \eta_{fin}) = \sum_{h=1}^H p_h(\lambda) b(y - c + d \log \lambda; \pi_h, \mu_h, \sigma_{1h}, \sigma_{2h}). 
$$

\cite{norets2010approximation} proves that continuous conditional densities for which i) the distribution of the covariate has bounded support, ii) the logarithm of the conditional density has finite expectation, and iii) the second moment of the outcome variable exists finite, such as the ones available in this work, can be approximated in the Kullback–Leibler divergence by finite mixtures of normal regressions in which means, variances, and mixing probabilities can depend on covariates:
$$
d_{KL} (f(\cdot | \cdot) , f_{fin}(\cdot|\cdot, c, d,\eta_{fin})) < \varepsilon,
$$
where $\eta_{fin} = \{\pi_h,\mu_h,\sigma_{1h},\sigma_{2h}\}_{h=1}^H$. Assumption 2, 4, and 5 of Section \ref{sec:meth} are needed to prove this result. 

From the convergence for finite mixtures, we will extend this result to infinite mixtures. 
A finite mixture can be characterised by Equation \eqref{eq:infinitemix}, and define the set of parameters for the first $H$ components as 
\begin{align*}
\eta^{1:H} &= \{V_h, \psi_h, \pi_h, \mu_h, \sigma_{1h}, \sigma_{2h}\}_{h=1}^H. 
\end{align*}
The infinite mixture can be characterised by a repetition of $\eta^{1:H}$, with renormalised weights, i.e.
\begin{align*}
\eta^{1:(HM)} &= \{\delta V_h, \psi_h, \pi_h, \mu_h, \sigma_{1h}, \sigma_{2h}\}_{h=1}^H \times \\
& \qquad \times \ldots \times \{\delta V_h, \psi_h, \pi_h, \mu_h, \sigma_{1h}, \sigma_{2h}\}_{h=1}^H 
\end{align*}
where $\delta \in (0,1)$. 
Let the rest of the parameters $\eta^{(HM+1):\infty}$ unconstrained. 

This is useful, because it allows us to apply Proposition 4.1 of \cite{norets2014posterior}: 
\begin{align*}
f(y | &\lambda, c, d, \eta^{1:(HM)}) =  \sum_{h=1}^{HM} \delta V_h w(\lambda,\psi_h) \prod_{\ell < h} (1-\delta V_{\ell} w(\lambda,\psi_{\ell})) \cdot \\
& \quad b(y - c + d \log \lambda ; \pi_h, \mu_h, \sigma_{1h}, \sigma_{2h}) \\
& >  f(y | \lambda, c, d, \eta_{fin}) (1 - \delta \max_{h=1,\ldots, H} V_h)^H \times \left( 1 - \left[ \prod_{\ell=1}^H (1-\delta V_{\ell} w(\lambda, \psi_h))\right]^M \right). 
\end{align*}

Then there exists $\delta > 0$ small enough and $M$ large enough such that 
$$
\log \left( \frac{f_{fin}(y | \lambda, c, d, \eta_{fin})}{f(y | \lambda, c, d, \eta^{1: (HM)})} \right) < \varepsilon.
$$

Next we want to show that, for any $\varepsilon>0$ and any $(c, d,\eta^{1:(HM)})$, there exists an open neighbourhood $\Upsilon$ such that for any $(\tilde{c},\tilde{d},\tilde{\eta}^{1:(HM)}) \in \Upsilon$ 
$$
\int \log \frac{f(y| \lambda, c, d, \eta^{1:(HM)})}{f(y | \lambda, \tilde{c},\tilde{d}, \tilde{\eta}^{1:(HM)})} dF_0 (y, \lambda) < \varepsilon. 
$$
To show this, we use the dominated convergence theorem. Consider a sequence $(c^{\ell}, d^{\ell}, \eta^{1:(HM), \ell})$ of parameter values converging to $(c, d,\eta^{1:(HM)})$ as $\ell \rightarrow \infty$; then
\begin{equation}
\log \frac{f(y | \lambda, c, d, \eta^{1:(HM)})}{p(y | \lambda, c^{\ell}, d^{\ell}, \eta^{1:(HM), \ell})} \rightarrow 1. 
\label{eq:proof_second}
\end{equation}

To apply the dominated convergence theorem, we need to show that there exists an integrable (with respect to $F_0$) upper bound of $|\log f(y | \lambda, c^{\ell}, d^{\ell}, \eta^{1:(HM), \ell})|$. Since every convergent sequence is bounded in a metric space and $(c^{\ell}, d^{\ell}, \eta^{1:(HM), \ell}) \rightarrow (c,d, \eta^{1:(HM)})$, this implies bounds on the parameters of the mixtures for $\ell$ large enough, i.e. $\mu_h^{\ell} \in (\mu_{l}, \mu_{u})$, $-\mu_h^{\ell} \frac{\pi_h^{\ell}}{1-\pi_h^{\ell}} \in (\mu_l, \mu_u)$, and $\sigma_{1h}^{\ell} \in (\sigma_{l}, \sigma_{u})$,  $\sigma_{2h}^{\ell} \in (\sigma_{l}, \sigma_{u})$. For $\ell$ large enough, $(c^{\ell},d^{\ell}) \in N_{\varepsilon}(c,d)$, and $(c^{\ell}-d^{\ell} \log \lambda) \in (\xi_l, \xi_u)$ with
\begin{align*}
\xi_l &= \inf_{(\tilde{c},\tilde{d}) \in N_{\varepsilon}, \lambda \in \Lambda} \tilde{c} - \tilde{d} \log \lambda \\
\xi_u &= \sup_{(\tilde{c},\tilde{d}) \in N_{\varepsilon}, \lambda \in \Lambda} \tilde{c} - \tilde{d} \log \lambda.
\end{align*}
Define $\nu_l = \mu_l + \xi_l$ and $\nu_u = \mu_u + \xi_u$. Then by Theorem 3.1 of \cite{norets2014posterior} 
\begin{align*}
\frac{\psi(0)}{\sigma_{l}} & \geq f(y|\lambda, c^{\ell},d^{\ell}, \eta^{1:(HM), \ell}) \geq \frac{1}{\sqrt{2\pi \sigma^2_u}} \left[ \mathbb{I}_{(-\infty,\nu_l)} \exp\left( - \frac{(y-\nu_u)^2}{2 \sigma^2_l}\right) + \right. \\
& \left. \mathbb{I}_{(\nu_l, \nu_u)} \exp\left( - \frac{(\nu_l-\nu_u)^2}{2 \sigma^2_l}\right) + \mathbb{I}_{(\nu_u,\infty)} \exp\left( - \frac{(y-\nu_l)^2}{2 \sigma^2_l}\right)\right].
\end{align*}
where $\psi(z)$ is a bounded, continuous, symmetric around zero, and monotone decreasing in $|z|$ probability density. 

The upper bound of this expression is a finite constant, hence it is integrable. The logarithm of the lower bound is integrable because of Assumption 3 of Theorem \ref{th:approx}.  

Finally, consider any $(\tilde{c},\tilde{d}, \tilde{\eta}) \in \Upsilon$, let $\eta^{\infty} = (\tilde{\eta}, \eta^{(HM+1):\infty})$ where $\eta^{(HM+1):\infty}$ represents the unrestricted part of the model. Then, for any $(y, \lambda)$
$$
\log \left( \frac{f(y | \lambda,\tilde{c},\tilde{d}, \tilde{\eta})}{f(y | \lambda, \tilde{c},\tilde{d}, \eta^{\infty})} \right) < 0. 
$$

Then, there exists $(c,d, \eta^{1:(HM)})$ and an open neighbourhood $\Upsilon$ of $(c,d, \eta^{1:(HM)})$ such that, for any $(\tilde{c},\tilde{d},\tilde{\eta}^{1:(HM)}) \in \Upsilon$ and any $\eta^{\infty} = (\tilde{\eta}^{1:(HM)}, \eta^{(HM+1):\infty})$, 
\begin{align*}
\int \log & \frac{f_0(y | \lambda)}{f(y | \lambda, c,d, \eta^{\infty})} dF_0(y, \lambda) = \\
&= \int \log \frac{f_0(y | \lambda)}{f_{fin}(y | \lambda, c,d, \eta_{fin})} dF_0(y, \lambda) + \int \log \frac{f_{fin}(y | \lambda, c,d, \eta_{fin})}{f(y | \lambda, c,d, \eta^{1:(HM)})} dF_0(y, \lambda) + \\
& \qquad \int \log \frac{f(y | \lambda, c,d, \eta^{1:(HM)})}{f(y | \lambda, \tilde{c},\tilde{d}, \tilde{\eta}^{1:(HM)})} dF_0(y, \lambda) + \int \log \frac{f(y | \lambda, \tilde{c},\tilde{d}, \tilde{\eta}^{1:(HM)})}{f(y | \lambda, \tilde{c},\tilde{d}, \eta^{\infty})} dF_0(y, \lambda) \\
& < \varepsilon + \varepsilon + \varepsilon + 0 < 3\varepsilon.
\end{align*} 

If $\eta^{1:(HM)}$ is in the support of the prior and the priors assigns a positive mass for some neighbourhood of any $\eta^{1:(HM)}$, then $\Pi(\Upsilon)> 0$ and $f_0 \in KL(\Pi)$. 

\newpage

\item[Title:] Proof of Theorem \ref{th:consistency}. 

We need to construct the necessary exponentially consistent tests. Define $\beta=(c,d)$ and $\beta_0=(c_0,d_0)$. Then 
\begin{align*}
H_0: & \; (f,\beta) = (f_0, \beta_0) \\
H_1: & \; (f,\beta) \in \mathcal{V} \times \{\parallel \beta-\beta_0\parallel <\rho \} 
\end{align*}
where $\mathcal{V} = supp(\Pi_{f_{u | \lambda}})$. We construct a finite set of alternative hypotheses such that their union is a superset of $H_1$. Define $\gamma \in \{-1,1\}$ and the quadrant $Q_{\gamma} = \{z \in \mathbb{R}^2: z_j\gamma_j > 0, \; \forall j=1,2\}$. Consider $\Delta > 0$ and define 
\begin{align*}
H_0: & \; (f,\beta) = (f_0, \beta_0) \\
H_1: & \; (f,\beta) \in \mathcal{V} \times \{ \beta: \parallel \beta-\beta_0\parallel \in Q_{\gamma}, \gamma_j(\beta_j-\beta_{0j}) > \Delta , \; \forall j=1,2\ \}. 
\end{align*}

For ease of notation, redefine 
\begin{equation*}
c - d \log \lambda = c-d \mathbb{E}_{F_0}[\log \lambda] + d(\log \lambda - \mathbb{E}_{F_0}[\log \lambda])
\end{equation*}
where $F_0$ is the joint distribution of $y$ and $\log \lambda$. 

Since the domain of $\log \lambda$ is bounded, we can find $\xi>0$ small enough that for some $\zeta > 0$
$$
F_0(Q_{\gamma,>}) \geq \zeta > 0
$$
where $Q_{\gamma,>} = \{\lambda \in \Lambda | (\log \lambda - \mathbb{E}[\log \lambda]) \in Q_{\gamma}, -(\log \lambda - \mathbb{E}(\log \lambda)) > \xi\}$, and 
$$
F_0(Q_{-\gamma,<}) \geq \zeta > 0
$$
where $Q_{-\gamma,<} = \{\lambda \in \Lambda | (\log \lambda - \mathbb{E}[\log \lambda]) \in Q_{-\gamma}, -(\log \lambda - \mathbb{E}(\log \lambda)) < -\xi\}$. 

For $(\beta-\beta_0) \in Q_{\gamma}$ consider
\begin{align*}
& c - c_0 - (d-d_0) \log \lambda = c - c_0 - (d-d_0)\mathbb{E}(\log \lambda) - (d-d_0)(\log \lambda - \mathbb{E}(\log \lambda)) \\
& Q_{\gamma, +} \text{ s.t. } c-c_0 - \mathbb{E}(\log \lambda)(d-d_0) \geq 0 \\
& Q_{\gamma, -} \text{ s.t. } c-c_0 - \mathbb{E}(\log \lambda)(d-d_0) < 0 \\
& Q_{\gamma, +} \cup Q_{\gamma,-} = Q_{\gamma}.
\end{align*}

\textbf{Step 1.} Construct a test for $(\beta-\beta_0) \in Q_{\gamma,+}$. First, let $\gamma_2 > 0$, and consider only $\lambda \in Q_{\gamma,>}$ so that $-(\log \lambda - \mathbb{E}(\log \lambda))\gamma_2 > \xi$. 

By construction
\begin{align*}
& - (\log \lambda - \mathbb{E}(\log \lambda)) (d-d_0) > \xi \Delta, \text{ and }\\
& F_0(Q_{\gamma,>}) \geq \zeta > 0
\end{align*}

\textbf{Step 2.} Construct a test for $(\beta-\beta_0) \in Q_{\gamma,-}$. Let $\gamma_2 < 0$, and consider only $\lambda \in Q_{\gamma,<}$, so that $-(\log \lambda - \mathbb{E}(\log \lambda)) \gamma_2 < -\xi$.  . Then, by construction
\begin{align*}
& - (\log \lambda - \mathbb{E}(\log \lambda))(d-d_0) < -\xi \Delta, \text{ and }\\
& F_0(Q_{-\gamma,<}) \geq \zeta > 0. 
\end{align*}

The Chebyshev's inequality can be used to construct a test for $H_{1+}$, where $H_1$ is restricted to $Q_{\gamma,+}$. We need two assumptions
\begin{enumerate}
\item $\mathbb{E}[y | \lambda] = c_o - d_0 \log \lambda$ almost surely in $F_0$
\item $f_0(y|\lambda)$ is continuous in $(y,\lambda)$ a.s. $F_0$. 
\end{enumerate}

Then there exists $M$ such that $M > \sup_{\lambda \in \Lambda} \mathbb{E}_{F_0} [(y - c + d_0 \log \lambda) | \lambda]$. 

For any $m$, define $K_{m,>} = \sum_{j=1}^m \mathbb{I}\{\lambda_j \in Q_{\gamma,>}\}$ and 
$$
T_{m,>} = \frac{1}{K_{m,>}} \sum_{j=1}^m \mathbb{I}\{\lambda_j \in Q_{\gamma,>}\}(y_j - c_0 + d_0 \log \lambda_j). 
$$

By the Chebychev's inequality, for $\varepsilon>0$,
$$
P_{f_0}(|T_{m,>} \geq \varepsilon|) \leq \frac{M^2}{K_{m,>} \varepsilon^2} \overset{m \rightarrow \infty}{\rightarrow} 0.
$$

Now, we need to show that $\inf_{(f,\beta) \in H_{1+}}P_{f_0}(|T_{m,>} \geq \varepsilon|) \rightarrow 1$ as $m \rightarrow \infty$. 

Remember that $-(\log \lambda - \mathbb{E}(\log \lambda))(d-d_0) > \xi \Delta$ for every $\beta \in H_{1+}$. Consider $\tilde{T}_{m,>} = \frac{1}{K_{m,>}} \sum_{j=1}^m \mathbb{I}\{\lambda_j \in Q_{\gamma,>}\}(y_j - c + d \lambda_j)$. By construction, if $|\tilde{T}_{n,>}| \leq \epsilon$ under $H_{1+}$, then $|T_{n,>}| \geq \epsilon$ because we have defined 
$$
y_j - c + d \log \lambda_j = y_j - c_0 + d_0 \log \lambda_j - [c - c_0 - \mathbb{E}[\log \lambda] (d - d_0) - (\lambda_j - \mathbb{E}(\log \lambda))(d-d_0)],
$$
and the term in the squared brackets is larger than $\xi \Delta$ for $\lambda_j \in Q_{\gamma,>}$ is $H_{1,+}$ is true. 

Then 
$$
P_{f,\beta}(|T_{m,>}| \geq \varepsilon) \geq P_{f,\beta}(|\tilde{T}_{m,>}| \leq \varepsilon) \geq 1 - \frac{L^2}{K_{m,>} \varepsilon^2} \overset{m\rightarrow \infty}{\rightarrow} 1
$$
for any $(f,\beta) \in H_{1+}$, and where $L > \underset{{\lambda \in \Lambda, f \in \mathcal{U}_{\delta}(f_{0,u|\lambda}), c\in \mathbb{R}, d \in (-1,1/2)}}{\sup} \mathbb{E}_{f,\beta}[(y-c+d\log \lambda)^2 | \lambda]$.

These facts can be used to construct an uniformly consistent sequence of tests 
$$
\phi_m(\lambda_m,Y_m) = \mathbb{I}\left\{|T_{m,>}| \geq \varepsilon \right\}. 
$$

By Proposition 4.4.1 of \cite{ghosh2003bayesian}, this implies the existence of exponentially consistent tests. The case where $H_{1-}$ and $(\beta-\beta_0) \in Q_{\gamma,-}$ can be obtained in a similar way. 

Choosing $\Delta > 0$ small enough, the union of these sets of alternative hypotheses will contain the set
$$
(f,\beta) \in \mathcal{V} \times \{\beta : \parallel \beta-\beta_0\parallel \geq \rho\}. 
$$

Given a weak neighbourhood of $\mathcal{U}_{\delta}(f_{0,u|\lambda})$ of conditional error densities, we construct exponentially consistent tests
\begin{align*}
& H_0: \; (f_{u|\lambda}, \beta) = (f_{0,u|\lambda},\beta_0) \\
& H_1: \; (f_{u|\lambda}, \beta) = \mathcal{U}_{\delta}^C(f_{0,u|\lambda}) \times \{\beta: \parallel \beta-\beta_0\parallel < \rho \}
\end{align*}
for $\rho>0$. 

Since $\mathcal{U}_{\delta}(f_{0,u|\lambda})$ is defined via some bounded uniformly continuous function $g$, let $\Delta_2$ be such that if $|u_1 - u_2| < \Delta_2$, then $| g(u_1, \lambda) - g(u_2,\lambda) | < \delta$. Then for $f_{u|\lambda} \in \mathcal{U}_{\delta}^C(f_{0,u|\lambda})$ and for any $\beta$ such that $\parallel \beta - \beta_0 \parallel < \rho$ with $\rho$ such that $|(c-c_0) - \lambda(d - d_0)| < \Delta_2$, define $f_{y|\lambda,\beta} = f_{u | \lambda}(u + c - d \log \lambda, \lambda)$. 

Then $f_{y|\lambda,\beta} \in \mathcal{U}_{\delta}^C(f_{0,y|\lambda})$ where $\mathcal{U}_{\delta}$ is a weak neighbourhood of $f_0(y|\lambda)$. Therefore, exponentially consistent tests always exist for Theorem 4.4.2 of \cite{ghosh2003bayesian}. If $\Delta_2$ and $\rho$ are taken small enough, the union would contain
$$
\{(f_{u|\lambda}) \in \mathcal{U}_{\delta}^C(f_{0,u|\lambda}) \times \{\beta: \parallel \beta - \beta_0 \parallel < \rho \}\}.
$$

Since $f_0 \in KL(\Pi)$ by Theorem 1, we can simply apply the results in \cite{schwartz1965bayes} to obtain that 
$$
\Pi(W^C | y_1, \ldots, y_m, \lambda_1, \ldots, \lambda_m) \rightarrow 0
$$
in $P_{F_0}^{\infty}$-probability for $\rho >0$, with $
W = \mathcal{U}_{\delta}(f_{0,u | \lambda}) \times \{\beta: \parallel \beta-\beta_0\parallel<\rho\}
$.

\newpage

\item[Title:] Results from the Simulation Study of Section \ref{sec:simu}

\begin{table}[h!]
\centering
\begin{footnotesize}
\begin{tabular}{c|c|c|c|c|c}
              & $\theta=0$, $\phi=0$ & $\theta=0.1,\phi=0.8$ &  $\theta=0.1,\phi=0.5$ & $\theta=0.5,\phi=0.1$ & $\theta=0.8$, $\phi=0.1$   \\
              \hline
\textbf{0.05} & 0.051         &   0.066 &  0.046 & 0.138 &   0.025  \\
              & (0.026,0.076) &  (0.002,0.135) &  (0.000,0.114) & (0.062,0.264) &  (0.000,0.092) \\
\textbf{0.10} & 0.101         &  0.117  &    0.097 & 0.160 &  0.079 \\
              & (0.078,0.125) &  (0.051,0.185)    &  (0.034,0.163) & (0.065,0.279) &  (0.012,0.150) \\
\textbf{0.15} & 0.152         &  0.163 &     0.146 & 0.175 & 0.129  \\
              & (0.129,0.182) &  (0.101,0.230)    &  (0.084,0.216) & (0.062,0.312) &  (0.067,0.194) \\
\textbf{0.20} & 0.200         &  0.216 & 0.195   &  0.192 & 0.177   \\
              & (0.177,0.227) &  (0.153,0.286)  &  (0.135,0.263) &  (0.088,0.321) &   (0.116,0.244) \\
\textbf{0.25} & 0.251         &  0.270  &  0.250 &  0.208 &  0.229  \\
              & (0.232,0.276) &  (0.205,0.339)    & (0.187,0.319)  & (0.102,0.328) &  (0.166,0.297) \\
\textbf{0.30} & 0.299         &  0.318 & 0.298 &  0.345 &  0.280  \\
              & (0.273,0.322) &  (0.255,0.385)   & (0.235,0.364) &  (0.206,0.453)  &  (0.216,0.350) \\
\textbf{0.35} & 0.349         &  0.373   & 0.352 & 0.388 &   0.327 \\
              & (0.330,0.371) &  (0.312,0.439)    & (0.288,0.416)  & (0.266,0.496)  &  (0.265,0.395) \\
\textbf{0.40} & 0.399          & 0.426 & 0.406   &   0.406   &   0.383 \\
              & (0.376,0.426)        &                         (0.363,0.495) & (0.343,0.474) &  (0.303,0.488) &   (0.320,0.450) \\
\textbf{0.45} & 0.451                & 0.477        &   0.459  &   0.481   &  0.437  \\
              & (0.431,0.474)        &                     (0.412,0.500)   &  (0.399,0.500) & (0.426,0.500)  &  (0.375,0.500) \\
\textbf{0.49} & 0.490                &  0.495            &   0.500  &    0.446   &   0.480 \\
              & (0.460,0.500)        &                         (0.450,0.500) & (0.437,0.500) & (0.291,0.496) & (0.414,0.500)
\end{tabular}
\end{footnotesize}
\caption{Average MAP estimates and relative 95\% credible intervals for the parameter $d$ of ARFIMA models obtained across 50 repetitions of the experiments.}
\label{tab:simu_mine}
\end{table}

\begin{table}[h!]
\centering
\begin{small}
\begin{tabular}{c|c|c|c|c|c}
              & $\theta=0$, $\phi=0$ & $\theta=0.1,\phi=0.8$ &  $\theta=0.1,\phi=0.5$ & $\theta=0.5,\phi=0.1$ & $\theta=0.8$, $\phi=0.1$   \\
              \hline
\textbf{0.05} & 0.065 & 0.568 & 0.249 & -0.131 & -0.447 \\
              & (0.024,0.120) & (0.508,0.629) & (0.192,0.310) & (-0.189,-0.071) & (-0.505,-0.391) \\
\textbf{0.10} & 0.114 & 0.618 & 0.299 & -0.081 & -0.398 \\
              & (0.075,0.171) & (0.556,0.680) & (0.240,0.362) & (-0.140,-0.019) & (-0.456,-0.339) \\
\textbf{0.15} & 0.165 & 0.668 & 0.349 & -0.031 & -0.349 \\
              & (0.125,0.222) & (0.606,0.730) & (0.290,0.414) & (-0.090,0.033) & (-0.406,-0.288) \\
\textbf{0.20} & 0.215 & 0.718 & 0.399 & 0.019 & -0.299 \\
              & (0.174,0.273) & (0.656,0.779) & (0.340,0.464) & (-0.039,0.084) & (-0.356,-0.236) \\
\textbf{0.25} & 0.265 & 0.768 & 0.450 & 0.069 & -0.249 \\
              & (0.222,0.323) & (0.708,0.829) & (0.390,0.514) & (0.012,0.135) & (-0.307,-0.185) \\
\textbf{0.30} & 0.315 & 0.818 & 0.500 & 0.119 & -0.199 \\
              & (0.269,0.374) & (0.758,0.879) & (0.441,0.563) & (0.063,0.183) & (-0.256,-0.133) \\
\textbf{0.35} & 0.365 & 0.866 & 0.545 & 0.170 & -0.150 \\
              & (0.316,0.426) & (0.808,0.929) & (0.492,0.612) & (0.114,0.238) & (-0.203,-0.082) \\
\textbf{0.40} & 0.414 & 0.916 & 0.599 & 0.219 & -0.100 \\
              & (0.363,0.478) & (0.859,0.979) & (0.542,0.661) & (0.164,0.288) & (-0.150,-0.029) \\
\textbf{0.45} & 0.466 & 0.965 & 0.650 & 0.270 & -0.049 \\
              & (0.413,0.529) & (0.908,1.024) & (0.591,0.711) & (0.213,0.336) & (-0.099,0.022) \\
\textbf{0.49} & 0.506 & 1.004 & 0.689 & 0.310 & -0.009 \\
              & (0.465,0.568) & (0.945,1.065) & (0.628,0.758) & (0.251,0.377) & (-0.061,0.062)
\end{tabular}
\end{small}
\caption{Average frequentist estimates and relative 95\% confidence intervals for the parameter $d$ of ARFIMA models obtained across 50 repetitions of the experiments; estimates are obtained with the \texttt{gph} function available in the \texttt{LongMemoryTS} R package \cite{leschinski2019longmemoryts}.}
\label{tab:simu_freq}
\end{table}

\begin{table}[h!]
\centering
\begin{small}
\begin{tabular}{c|c|c|c|c|c}
              & $\theta=0$, $\phi=0$ & $\theta=0.1,\phi=0.8$ &  $\theta=0.1,\phi=0.5$ & $\theta=0.5,\phi=0.1$ & $\theta=0.8$, $\phi=0.1$   \\
              \hline
\textbf{0.05} & 0.058 & -0.030 & 0.052 & 0.072 & 0.003 \\
              & (0.017,0.092) & (-0.262,0.178) & (-0.096,0.434) & (-0.018,0.473) & (-0.154,0.278) \\
\textbf{0.10} & 0.108 & 0.018 & 0.109 & 0.106 & 0.055 \\
              & (0.067,0.142) & (-0.212,0.221) & (-0.051,0.499) & (0.014,0.474) & (-0.109,0.321) \\
\textbf{0.15} & 0.158 & 0.065 & 0.157 & 0.181 & 0.107 \\
              & (0.117,0.192) & (-0.160,0.266) & (-0.006,0.302) & (0.010,0.420) & (-0.064,0.360) \\
\textbf{0.20} & 0.207 & 0.119 & 0.204 & 0.226 & 0.157 \\
              & (0.167,0.242) & (-0.107,0.311) & (0.041,0.322) & (0.008,0.399) & (-0.018,0.393) \\
\textbf{0.25} & 0.257 & 0.170 & 0.253 & 0.246 & 0.120 \\
              & (0.216,0.292) & (-0.054,0.356) & (0.090,0.368) & (0.006,0.401) & (0.028,0.395) \\
\textbf{0.30} & 0.307 & 0.222 & 0.302 & 0.284 & 0.242 \\
              & (0.266,0.342) & (-0.008,0.398) & (0.142,0.411) & (0.007,0.409) & (0.078,0.415) \\
\textbf{0.35} & 0.356 & 0.271 & 0.350 & 0.333 & 0.285 \\
              & (0.316,0.391) & (0.037,0.429) & (0.201,0.448) & (0.009,0.439) & (0.133,0.433) \\
\textbf{0.40} & 0.406 & 0.309 & 0.392 & 0.346 & 0.339 \\
              & (0.367,0.440) & (0.084,0.453) & (0.263,0.471) & (0.010,0.449) & (0.197,0.447) \\
\textbf{0.45} & 0.454 & 0.354 & 0.428 & 0.378 & 0.383 \\
              & (0.415,0.482) & (0.133,0.468) & (0.318,0.482) & (0.008,0.488) & (0.257,0.464) \\
\textbf{0.49} & 0.485 & 0.381 & 0.451 & 0.406 & 0.411 \\
              & (0.453,0.496) & (0.172,0.474) & (0.355,0.486) & (0.017,0.486) & (0.293,0.475) 
\end{tabular}
\end{small}
\caption{Average MAP estimates and relative 95\% confidence intervals for the parameter $d$ of ARFIMA models obtained across 50 repetitions of the experiments when a parametric model is applied and parameter estimation is obtained through MCMC.}
\label{tab:simu_param}
\end{table}

\begin{table}[h!]
\centering
\begin{scriptsize}
\begin{tabular}{c|c|c|c|c|c|c}
              & & $\theta=0$, $\phi=0$ & $\theta=0.1,\phi=0.8$ &  $\theta=0.1,\phi=0.5$ & $\theta=0.5,\phi=0.1$ & $\theta=0.8$, $\phi=0.1$   \\
              \hline
\textbf{0.05} & Bayes & 0.95 & 0.94 & 0.94 & 0.96 & 0.88 \\
 & Freq & 0.95 & 0.77 & 0.77 & 0.70 & 0.72 \\   
 & Param & 0.95 & 0.52 & 0.90 & 0.90 & 0.66 \\ 
\textbf{0.10} & Bayes & 0.95 & 0.94 & 0.94 & 0.94 & 0.88 \\  
 & Freq & 0.95 & 0.80 & 0.82 & 0.80 & 0.76 \\
 & Param & 0.95 & 0.50 & 0.86 & 0.90 & 0.68 \\ 
\textbf{0.15} & Bayes & 0.95 & 0.94 & 0.92 & 0.94 & 0.90 \\  
 & Freq & 0.95 & 0.78 & 0.73 & 0.70 & 0.75 \\
 & Param & 0.95 & 0.54 & 0.92 & 0.92 & 0.68 \\ 
\textbf{0.20} & Bayes & 0.95 & 0.92 & 0.94 & 0.94 & 0.92 \\  
 & Freq & 0.95 & 0.80 & 0.70 & 0.60 & 0.73 \\
 & Param & 0.95 & 0.54 & 0.94 & 0.92 & 0.66 \\ 
\textbf{0.25} & Bayes & 0.95 & 0.92 & 0.94 & 0.96 & 0.94 \\  
 & Freq & 0.95 & 0.79 & 0.70 & 0.60 & 0.81 \\
 & Param & 0.95 & 0.54 & 0.94 & 0.94 & 0.66 \\ 
\textbf{0.30} & Bayes & 0.95 & 0.92 & 0.94 & 0.96 & 0.94 \\  
 & Freq & 0.95 & 0.79 & 0.75 & 0.60 & 0.77 \\
 & Param & 0.95 & 0.52 & 0.94 & 0.94 & 0.68 \\ 
\textbf{0.35} & Bayes & 0.95 & 0.90 & 0.94 & 0.96 & 0.92 \\  
 & Freq & 0.95 & 0.78 & 0.75 & 0.65 & 0.78 \\
 & Param & 0.95 & 0.52 & 0.94 & 0.96 & 0.70 \\ 
\textbf{0.40} & Bayes & 0.95 & 0.90 & 0.94 & 0.96 & 0.94 \\  
 & Freq & 0.95 & 0.76 & 0.79 & 0.60 & 0.76 \\
 & Param & 0.95 & 0.50 & 0.96 & 1.00 & 0.68 \\ 
\textbf{0.45} & Bayes & 0.95 & 0.92 & 0.96 & 0.98 & 0.92 \\  
 & Freq & 0.95 & 0.76 & 0.74 & 0.50 & 0.74 \\
 & Param & 0.95 & 0.46 & 0.98 & 1.00 & 0.60 \\ 
\textbf{0.49} & Bayes & 0.95 & 0.92 & 0.96 & 0.98 & 0.92 \\  
 & Freq & 0.95 & 0.71 & 0.72 & 0.45 & 0.76 \\
 & Param & 0.90 & 0.40 & 0.98 & 1.00 & 0.52 \\ 
\end{tabular}
\end{scriptsize}
\caption{Coverage of the three compared approaches (Bayesian semiparametric, frequentist and Bayesian parametric) based on 95\% credible or confidence intervals centred at the posterior mean or least squares estimator of $d$.}
\label{tab:simu_coverage}
\end{table}

\end{description}

\FloatBarrier

\end{document}